\documentclass{article}[12pt]
\usepackage{amsmath,amssymb,amsfonts}
\usepackage{graphics, subfigure,color} 

\usepackage{latexsym}
\usepackage[dvips]{graphicx}
\usepackage{epsfig}

\usepackage{hyperref}

\numberwithin{equation}{section}

\newtheorem{prop}{Proposition}[section]

\newtheorem{lemma}[prop]{Lemma}

\newtheorem{cor}[prop]{Corollary}

\newtheorem{theorem}[prop]{Theorem}

\newcommand{\cP}{{\cal{P}}}
\newcommand{\cF}{{\cal{F}}}

\newcommand{\cB}{{\cal B}}
\newcommand{\cT}{{\cal T}}
\newcommand{\cS}{{\cal S}}
\newcommand{\cJ}{{\cal J}}
\newcommand{\cL}{{\cal L}}

\newcommand{\bbone}{{\bf 1}}

\newcommand{\bee}{\begin{equation}}
\newcommand{\ee}{\end{equation}}
\newcommand{\beq}{\begin{eqnarray}}
\newcommand{\eeq}{\end{eqnarray}}
\newcommand{\bqa}{\begin{eqnarray}}
\newcommand{\eqa}{\end{eqnarray}}
\newcommand{\bea}{\begin{eqnarray}}
\newcommand{\eea}{\end{eqnarray}}
\newcommand{\beann}{\begin{eqnarray*}}
\newcommand{\eeann}{\end{eqnarray*}}

\newcommand{\prf}{{\noindent \bf Proof\; \; }}
\newcommand{\qed}{{\hfill $\Box$}}

\makeatletter

\begin{document}

\title{The Multiscale Loop Vertex Expansion}

\author{Razvan Gurau\footnote{rgurau@cpht.polytechnique.fr; 
CPHT - UMR 7644, CNRS, \'Ecole Polytechnique, 91128 Palaiseau cedex, France, EU 
and Perimeter Institute for Theoretical Physics, 31 Caroline St. N, N2L 2Y5, Waterloo,  ON, Canada. } , 
Vincent Rivasseau\footnote{vincent.rivasseau@th.u-psud.fr; 
LPT - UMR 8627, CNRS, Universit\'e Paris 11, 91405 Orsay Cedex, France, EU 
and Perimeter Institute for Theoretical Physics, 31 Caroline St. N, N2L 2Y5, Waterloo, ON, Canada. }
}
\maketitle
\begin{abstract}
The loop vertex expansion (LVE) is a constructive technique which uses only canonical combinatorial tools
and no space-time dependent lattices. It works for quantum field theories without renormalization.
Renormalization requires scale analysis. In this paper we provide an enlarged formalism
which we call the multiscale loop vertex expansion (MLVE). We test it on what is probably the simplest
quantum field theory which requires some kind of renormalization, namely a combinatorial model of the vector type
with quartic interaction and a propagator which mimicks the power counting of $\phi^4_2$.
An ordinary LVE would fail to treat even this simplest superrenormalizable model, but we show how to 
perform the ultraviolet limit and prove its analyticity in the Borel summability domain of the model with the MLVE.

\end{abstract}

\section{Introduction}
\label{intro}

Constructive field theory \cite{GJ} is a set of techniques to resum perturbative quantum field theory 
and obtain a rigorous definition of quantities such as the Schwinger functions of interacting renormalized models.
Although superrenormalizable and even just renormalizable
models have been treated in the past \cite{Riv}, it has the reputation of being a difficult technical subject. 
This reputation is mostly due to the complicated formalism of iterated cluster and Mayer expansions, also known
as the phase space expansion. In spite of many great and commendable efforts to improve the presentation of this undoubtedly powerful formalism,
it has remained quite confidential. Part of the problem lies within the use of non-canonical spatial tools, namely the space-time lattices. 
They break the natural rotation invariance of the theory and have certainly hindered the development of constructive gauge theories, of constructive
field theory in curved space time and of constructive models of random space-times.

The (single scale) loop vertex expansion (LVE) is a formalism designed to improve on traditional constructive tools. 
For ultraviolet convergent theories or in a single renormalization group slice it can be viewed 
as a reorganization of the perturbative series \cite{Rivasseau:2013ova}. It combines an intermediate field functional integral representation 
with a forest formula \cite{BK,AR1} and a replica trick. Remark that such canonical combinatorial tools are independent of the space-time geometry.
It allows the computation of connected Schwinger functions as convergent sums indexed by spanning trees of arbitrary size $n$ rather than divergent sums indexed by Feynman graphs. Indeed trees proliferate not as fast as general Feynman graphs when their number $n$ of vertices increases.
Initially introduced to analyze \emph{matrix} models with quartic interactions \cite{Rivasseau:2007fr}, the LVE can been
extended to other stable polynomial interactions \cite{RW}, and has been shown compatible with
direct space decay estimates \cite{MR1}.

Recently interest has risen in quantum gravity for the study of combinatorial quantum field theories which do not refer to any
particular space-time structure but are formulated on random tensors of rank $D$ \cite{Gurau:2011xp,Rivasseau:2013uca}.
Such models are characterized by a $U(N)^{\otimes D}$ invariance. Their large $N$ regime relies on a different $1/N$ expansion for $D=1$ (vector models),
$D=2$ (matrix models) and $D \ge 3$ (proper tensor models). When $D=2$, matrix models 
are well identified with a discretized version of two-dimensional quantum gravity \cite{Di Francesco:1993nw}. The LVE has been 
shown particularly well adapted to the analysis of random \emph{tensor} models of rank $D \ge 3$, in which
it computes in a natural way the $1/N$ expansion \cite{sefu1,Gurau:2013pca}.
Breaking this $U(N)^{\otimes D}$ invariance at the propagator level creates combinatorial quantum field theories with 
a full-fledged renormalization group flow \cite{BenGeloun:2011rc,Carrozza:2012uv,Carrozza:2013wda,Geloun:2013saa}. For just renormalizable models
with quartic interactions it is interesting to notice that this flow is non-asymptotically free in the vector case, asymptotically safe in the matrix 
case \cite{Disertori:2006nq} and asymptotically free in the proper tensor case \cite{BenGeloun:2012pu,BenGeloun:2012yk}.

Probably the simplest example of a combinatorial field theory which requires some renormalization is
a model of the vector type ($D=1$) with quartic interaction and  a covariance $\delta_{pq}/p$ between components $p$ and $q$. 
Its power counting is similar to the one of the $\phi^4_2$ model \cite{Simon} but it lacks the technicalities due
to space-time decomposition. Its renormalized interaction is still positive at $\lambda >0$, hence it does not
even have the problem called Nelson's bound \cite{nelson}. However to treat constructively 
even this extremely simple theory requires a multiscale analysis in the sense of the renormalization group. 

In this short pedagogical paper we provide a multiscale version of the loop vertex expansion by combining the intermediate field representation
of the ordinary LVE with a multiscale analysis and \emph{two} successive forests formulas. The first formula acts as in the ordinary LVE
on the intermediate Bosonic fields. The second acts on Fermionic fields, implementing a kind of Mayer expansion between the blocks 
created by the first expansion\footnote{Remark that although Fermionic forest formulas are rather standard in constructive field theory \cite{abdesselam,disertori}, 
we do not know if they were used before this paper to implement Mayer expansions.}.
Our toy model example hopefully will help disentangle the combinatorial core of this MLVE, which we treat in great detail, 
from other unessential aspects.

The plan of this paper is as follows: section 2 contains the model, the MLVE and the statement of the main convergence and analyticity theorem
which is then proved in section 3. Appendices 1 and 2 collect for completeness standard facts about the forest formula and the Mayer expansion.

\section{The Model}

Consider a pair of conjugate vector fields $\{\phi_p \}, \{\bar \phi_p \},  p=1 ,\cdots , N $, 
with $\frac{\lambda^2 }{2}(\bar \phi \cdot \phi )^2$ bare interaction, where $ (\bar \phi \cdot \phi ) \equiv \sum_{p=1}^N \bar \phi_p \phi_p $ . 
The Gaussian measure $d\eta (\bar \phi, \phi)$  is chosen to break the $U(N)$ invariance of the theory. It has
diagonal covariance (or propagator) which decreases as the inverse power of the field index:
\[
 d\eta(\bar \phi, \phi) = \Bigl(  \prod_{p=1}^N p  \frac{ d\bar \phi_p d\phi_p }{2\pi \imath} \Bigr)  e^{-\sum_{p=1}^N  p \;  \bar \phi_p \phi_p } \; ,
\qquad  \int  d\eta (\bar \phi, \phi) \; \bar\phi_p \phi_{q}= \frac{\delta_{pq}}{p} \;.
\]
This propagator renders the perturbative amplitudes of the model finite in the $N \to \infty$ limit, except for a mild 
divergence of self-loops which yield a logarithmically divergent sum
$
L_N = \sum_{p=1}^N \frac{1}{p}   \simeq \log N
$.
These divergences are easily renormalized by using a vector-Wick-ordered $\phi^4$ interaction, 
namely $\frac{1}{2}[\lambda (\bar \phi \cdot \phi -L_N)]^2$. Remark that this interaction 
(contrary to the $\phi^4_2$ case) remains positive for $\lambda$ real at all values of $(\bar\phi, \phi)$.
The renormalized partition function of the model is:
\bea\label{eq:partitionfunction} 
Z(\lambda, N) =  \int  d\eta (\bar \phi, \phi ) \; \;  e^{- \frac{\lambda^2}{2} (\bar \phi \cdot \phi -L_N)^2 }.
\eea
The intermediate field representation decomposes the quartic interaction using an intermediate scalar field $   \sigma  $:
\[
e^{- \frac{\lambda^2}{2} (\bar \phi \cdot \phi -L_N)^2 }    =  \int d\nu  (\sigma) \; 
e^{ \imath \lambda  \sigma   (\bar \phi \cdot \phi -L_N) } \; ,
\]
where $d\nu(\sigma) = \frac{1}{\sqrt{2\pi}} e^{-\frac{\sigma^2}{2} }$ is the standard 
Gaussian measure with covariance 1. Integrating over the initial fields $(\bar \phi_p, \phi_p)$ leads to:
\[
Z(\lambda, N)  =   \int d\nu  (\sigma) \; 
     \prod_{p=1}^N \frac{1}{1- \imath \frac{\lambda \sigma}{p}} e^{-\imath \frac{\lambda \sigma}{p} } 
    =  \int d\nu  (\sigma) \;  e^{- \sum_{p=1}^N \log_2 \bigl(1 - \imath \frac{\lambda  \sigma }{p } \bigr) } \; ,
\]
where $\log_2 (1-x) \equiv x+ \log (1-x) = O(x^2)$.

Performing a single scale standard LVE expansion on this functional integral, even for this most
simple of all renormalizable models, already runs into trouble. The LVE expresses $\log Z(\lambda, N)$ a sum over trees, 
but there is no simple way to remove the logarithmic divergence of all leaves of the tree without generating many intermediate fields 
in numerators which, when integrated through the Gaussian measure, would create an apparent divergence of the series. This will be 
explained in more detail below. Here we proceed differently.

We fix an integer $M>1$ and define the $j$-th \emph{slice}, as made of the indices 
$p \in I_j \equiv [M^{j-1},M^{j} -1]$. The ultraviolet cutoff $N$ is chosen as $N =M^{j_{max}}-1$, with $j_{\max}$ an integer.
We can also fix an infrared cutoff $j_{\text{min}}$. Hence there are $j_{\max}-j_{\min}$ 
slices in the theory, and the ultraviolet limit corresponds to the limit $j_{\max} \to \infty$. 
The intermediate field representation writes:
\bea\label{eq:partitionfunctionintfield}
Z(\lambda, N) =  \int d\nu  (\sigma) \; \prod_{j =j_{\min}}^{j_{\max}}     e^{- V_j} \; , \quad 
V_{j} =\sum_{p \in I_{j}} \log_2 \Bigl(1 - \imath \frac{ \lambda \sigma}{p}  \Bigr) \; ,
\eea
and we note that the interaction is now factorized over the set of slices $\cS = [j_{\min}, \cdots j_{\max}]$.
This factorization of the interaction can be encoded into an integral over Grassmann numbers. Indeed,
\[
   a = \int d\bar \chi d\chi \; e^{-\bar \chi a \chi} = \int d\mu (\bar \chi ,\chi ) \; e^{- \bar \chi (a-1) \chi}  
\]
where $d \mu(\bar \chi ,\chi ) = d\bar \chi d\chi \; e^{-\bar \chi \chi}$ is the standard normalized Grassmann Gaussian measure with covariance 1. Hence,
denoting $W_j(\sigma) = e^{-V_{j}}-1$, we can rewrite:
\[
Z(\lambda, N) =  \int d\nu  (\sigma) \; \Bigl( \prod_{j = j_{\min}}^{j_{\max} } d\mu (\bar \chi_j , \chi_j) \Bigr) \; 
    e^{ - \sum_{j = j_{\min}}^{j_{\max}}   \bar \chi_j  W_j   (  \sigma)   \chi_j } .
\]
Below we will use the fact that Gaussian integrals can be represented as derivative operators. 
The advantage of using this representation is that the covariances of the Gaussian measures are explicit.
For instance the partition function can be written as:
\[
  Z(\lambda, N) = \Bigl[ e^{\frac{1}{2} \frac{\partial}{\partial \sigma}\frac{\partial}{\partial \sigma} 
   +  \sum_{j=j_{\min}}^{j_{\max}} \frac{\partial}{\partial \bar \chi_j } \frac{\partial}{\partial \chi_j } } \;\;
   \prod_{j =j_{\min}}^{j_{\max}}   e^{ - \bar \chi_j  W_j   ( \sigma)   \chi_j } \Bigr]_{\sigma,\bar \chi_j,\chi_j =0 } \; .
\]

For any finite set $\cS$, let us denote $\bbone_{\cS}$ the $\vert \cS \vert $ by $\vert \cS \vert$ matrix 
with coefficients 1 everywhere and $\mathbb{I}_{\cS}$ the $\vert \cS \vert $ by $\vert \cS \vert$ identity matrix.
We introduce slice replicas for the Bosonic fields, that is we rewrite the partition function as:
\[
 Z(\lambda, N) =  \int d\nu_\cS \; e^{- W} \; ,
\quad d\nu_{\cS}  = d\nu_{\bbone_\cS}  ( \{ \sigma_j\}  ) \;  d\mu_{\mathbb{I}_\cS} (\{\bar \chi_j , \chi_j\})   \; ,
 \quad W = \sum_{j =j_{\min}}^{j_{\max}}   \bar \chi_j  W_j   ( \sigma_j)   \chi_j \; , 
\]
or equivalently:
\[
 Z(\lambda, N) = \Bigl[ e^{\frac{1}{2} \sum_{j,k=j_{\min}}^{j_{\max } } \frac{\partial}{\partial \sigma_j}\frac{\partial}{\partial \sigma_k} 
   +  \sum_{j=j_{\min}}^{j_{\text{max}}} \frac{\partial}{\partial \bar \chi_j } \frac{\partial}{\partial \chi_j } } \;\;
    e^{ - \sum_{j =j_{\min}}^{j_{\max}}   \bar \chi_j  W_j   ( \sigma_j)   \chi_j } \Bigr]_{\sigma_j,\bar \chi_j ,\chi_j =0 } \; .
\]
This is the starting point for the MLVE. The first step is to expand to infinity the exponential of the interaction:
\[
Z(\lambda, N) =  \sum_{n=0}^\infty \frac{1}{n!}\int d\nu_{\cS}  \; (-W)^n \; .
\]
The second step is to introduce replica Bosonic fields for all the vertices in $V = \{1, \cdots , n\}$: 
\[
Z(\lambda, N) = \sum_{n=0}^\infty \frac{1}{n!}  \int d\nu_{\cS,V} \;  \prod_{a=1}^n  (-W_a) \; ,
\]
where the $a$-th vertex $W_a$ has now its own (replicated) Bosonic fields $\sigma_j^a$ and the replica measure is completely degenerated:
\[
d\nu_{\cS,V}  = d\nu_{\bbone_\cS \otimes \bbone_V} (\{  \sigma^a_j\}) \;  d\mu_{\mathbb{I}_\cS  } (\{\bar \chi_j , \chi_j\})   \; ,\quad 
W_a =  \sum_{j =j_{\min}}^{j_{\max}}   \bar \chi_j W_j   (  \sigma^a_j )  \chi_j \; .
\]
Note that, very importantly, we have not introduced any vertex replicas for the Fermionic fields. Expressing the Gaussian integral as a derivative
operator we write equivalently:
\[
  Z(\lambda, N) = \sum_{n=0}^\infty \frac{1}{n!} \;\;
  \Bigl[ e^{\frac{1}{2} \sum_{a,b=1}^n\sum_{j,k=j_{\min}}^{j_{\max}} \frac{\partial}{\partial \sigma^a_j}\frac{\partial}{\partial \sigma^b_k} 
   +  \sum_{j=j_{\min}}^{j_{\max}} \frac{\partial}{\partial \bar \chi_j  } \frac{\partial}{\partial \chi_j  } } 
  \; \prod_{a=1}^n \Bigl( -  \sum_{j =j_{\min}}^{j_{\max}}   \bar \chi_j W_j   (  \sigma^a_j )  \chi_j \Bigr) \Bigr]_{\sigma_j^a,\chi_j \bar\chi_j =0}
   \; .
\]

The obstacle to factorize this integral over vertices lies now in the Bosonic degenerate blocks $\bbone_V$ and 
in the Fermionic fields (which couple the vertices $W_a$). In order to deal with this we will apply {\it two} successive
forest formulas. First, in order to disentangle the block $\bbone_V$ in the measure $d \nu$ 
we introduce the coupling parameters $x_{ab}=x_{ba}, x_{aa}=1$ between the vertex Bosonic replicas:
\[
 Z(\lambda, N) = \sum_{n=0}^\infty \frac{1}{n!} 
  \Bigl[ e^{\frac{1}{2} \sum_{a,b=1}^n x_{ab} \sum_{j,k=j_{\min}}^{j_{\max}} \frac{\partial}{\partial \sigma^a_j}\frac{\partial}{\partial \sigma^b_k} 
   +  \sum_{j=j_{\min}}^{j_{\max}} \frac{\partial}{\partial \bar \chi_j  } \frac{\partial}{\partial \chi_j } } \; 
   \prod_{a=1}^n \Bigl( -  \sum_{j =j_{\min}}^{j_{\max}}   \bar \chi_j W_j   (  \sigma^a_j )  \chi_j \Bigr) 
   \Bigr]_{ \genfrac{}{}{0pt}{}{ \sigma, \chi,  \bar\chi  =0}{x_{ab}=1 }} \;,
\]
and apply the forest formula (see appendix \ref{app:forest}). We denote $\cF_B$ a Bosonic forest with $n$ vertices labelled $\{1,\dots n\}$, 
$\ell_{B}$ a generic edge of the forest and $a(\ell_B), b(\ell_B)$ the end vertices of $\ell_B$. The result of the first forest formula is:
\beann
&&  Z(\lambda, N) = \sum_{n=0}^\infty \frac{1}{n!} \sum_{\cF_{B}} \int_{0}^1 \Bigl( \prod_{\ell_B\in \cF_B } dw_{\ell_B} \Bigr)
\; \;  \Bigg[  e^{\frac{1}{2} \sum_{a,b=1}^n X_{ab}(w_{\ell_B}) \sum_{j,k=j_{\min}}^{j_{\max}}  
    \frac{\partial}{\partial \sigma^a_j}\frac{\partial}{\partial \sigma^b_k} 
   +  \sum_{j=j_{\min}}^{j_{\max}} \frac{\partial}{\partial \bar \chi_j  } \frac{\partial}{\partial \chi_j  } }
\crcr
&& \qquad \qquad \qquad \times 
   \prod_{\ell_B \in \cF_{B}} \Bigl( \sum_{j,k=j_{\min}}^{j_{\max}}  
   \frac{\partial}{\partial \sigma^{a(\ell_B)}_j}\frac{\partial}{\partial \sigma^{b(\ell_B)}_k}  \Bigr) \;
 \prod_{a=1}^n \Bigl( -  \sum_{j =j_{\min}}^{j_{\max}}   \bar \chi_j W_j   (  \sigma^a_j )  \chi_j \Bigr) \Bigg]_{\sigma ,\chi , \bar\chi =0 } \; ,
\eeann
where $X_{ab}(w_{\ell_B})$ is the infimum over the parameters $w_{\ell_B}$ in the unique path
in the forest $\cF_B$ connecting $a$ and $b$, and the infimum is set to 
$1$ if $a=b$ and to zero if $a$ and $b$ are not connected by the forest.

The forest $\cF_B$ partitions the set of vertices into blocks $\cB$ corresponding to its trees. Remark that the blocks can be singletons 
(corresponding to the trees with no edges in $\cF_B$). We denote $a\in \cB$ if the vertex $a$ belongs to a Bosonic block $\cB$. A vertex
belongs to a unique Bosonic block.
Contracting every Bosonic block to an ``effective vertex'' we obtain a graph which we denote $\{n\}/\cF_B$.
We introduce replica Fermionic fields $\chi^{\cB}_j$ for the blocks of $\cF_B$ (i.e. for the effective vertices 
of $\{n\}/\cF_B$) and replica coupling parameters $y_{\cB\cB'}=y_{\cB'\cB}$. 
Applying (a second time) the forest formula, this time for the $y$'s, leads to a set of Fermionic edges $\cL_F$ forming a forest 
in $\{n\}/\cF_B$ (hence connecting Bosonic blocks). We denote $L_{F} $ a generic Fermionic edge connecting blocks and $\cB(L_F), \cB'(L_F) $ 
the end blocks of the Fermionic edge $L_F$. We obtain: 
\bea\label{eq:intermediate}
&&  Z(\lambda, N) = \sum_{n=0}^\infty \frac{1}{n!} \sum_{\cF_{B}} \sum_{\cL_F}\int_{0}^1 \prod_{\ell_B\in \cF_B } dw_{\ell_B} 
\prod_{L_F\in \cL_F } dw_{L_F} \crcr
&& \quad  \quad \times \Bigg[  e^{\frac{1}{2} \sum_{a,b=1}^n X_{ab}(w_{\ell_B }) \sum_{j,k=j_{\min}}^{j_{\max }  } 
\frac{\partial}{\partial \sigma^a_j}\frac{\partial}{\partial \sigma^b_k} 
   +  \sum_{\cB,\cB'} Y_{\cB\cB'}(w_{\ell_F})\sum_{j=j_{\min}}^{j_{\max}}
   \frac{\partial}{\partial \bar \chi_j^{\cB} } \frac{\partial}{\partial \chi_j^{\cB'} } } \crcr
&& \quad \quad \times \prod_{\ell_B \in \cF_{B}} \Bigl( \sum_{j,k=j_{\min}}^{j_{\max }}  \frac{\partial}{\partial \sigma^{a(\ell_B)}_j}
 \frac{\partial}{\partial \sigma^{b(\ell_B)}_k}  \Bigr)
   \prod_{L_F \in \cL_F} 
   \Bigg(\sum_{j=j_{\min} }^{j_{\max}} \Bigl( \frac{\partial}{\partial \bar \chi_j^{\cB(L_F)} } \frac{\partial}{\partial \chi_j^{\cB'(L_F)} }  
   +   \frac{\partial}{\partial \bar  \chi_j^{\cB'(L_F)} } \frac{\partial}{\partial  \chi_j^{\cB(L_F)}} \Bigr)\Bigg)
   \crcr
&& \quad \qquad \quad \times  \prod_{\cB}  \prod_{a\in \cB} \Bigl( -  \sum_{j =j_{\min}}^{j_{\max }}   \bar \chi^{\cB }_j W_j   (  \sigma^a_j )  
\chi^{\cB }_j \Bigr)   \Bigg]_{\sigma ,\chi , \bar\chi =0 } \; ,
\eea 
where $Y_{\cB\cB'}(w_{\ell_F}) $ is the infimum over $w_{\ell_F}$ in the unique path in $\cL_{F}$ connecting $\cB$ and $\cB'$ and the infimum is 
set to $1$ if $\cB= \cB'$ and to zero if $\cB$ and $\cB'$ are not connected by $\cL_F$. Note that the Fermionic edges are oriented. 
Expanding the sums over $j$ in the last line of eq. \eqref{eq:intermediate} we obtain a sum over slice assignments $j_a$ to the vertices $a$, 
and taking into account that $\partial_{\sigma^a_j} W(\sigma^{a}_{j_a}) = \delta_{jj_a} \partial_{\sigma^a_{j_a} } W(\sigma^{a}_{j_a}) $
we obtain:
\beann
 &&  Z(\lambda, N) = \sum_{n=0}^\infty \frac{1}{n!} \sum_{\cF_{B}} \sum_{\cL_F} \;\sum_{j_1=j_{\min}}^{j_{\max }} 
  \dots \sum_{j_n=j_{\min}}^{j_{\max } }
 \; \int_{0}^1 \prod_{\ell_B\in \cF_B } dw_{\ell_B} 
\prod_{\ell_F\in \cL_F } dw_{L_F} \crcr
&& \quad \quad  \times \Bigg[  e^{\frac{1}{2} \sum_{a,b=1}^n X_{ab}(w_{\ell_B }) \frac{\partial}{\partial \sigma^a_{j_a} }\frac{\partial}{\partial \sigma^b_{j_b}} 
   +  \sum_{\cB,\cB'} Y_{\cB\cB'}(w_{\ell_F})\sum_{j=j_{\min}}^{j_{\max}} 
   \frac{\partial}{\partial \bar \chi_j^{\cB} } \frac{\partial}{\partial \chi_j^{\cB'} } } \crcr
&& \quad \quad \times \prod_{\ell_B \in \cF_{B}} \Bigl( \frac{\partial}{\partial \sigma_{j_{ a(\ell_B) }}^{a(\ell_B)} }
 \frac{\partial}{\partial \sigma^{b(\ell_B)}_{j_{  b(\ell_B) }}}  \Bigr)
   \prod_{L_F \in \cL_F} 
   \Bigg(\sum_{j=j_{\min}}^{j_{\max}} \Bigl( \frac{\partial}{\partial \bar \chi_j^{\cB(L_F)} } \frac{\partial}{\partial \chi_j^{\cB'(L_F)} }  
   +   \frac{\partial}{\partial \bar  \chi_j^{\cB'(L_F)} } \frac{\partial}{\partial  \chi_j^{\cB(L_F)}} \Bigr)\Bigg)
   \crcr
&& \quad \qquad \quad \times  \prod_{\cB}  \prod_{a\in \cB} \Bigl( -    \bar \chi^{\cB }_{j_a} W_{j_a}   (  \sigma^a_{j_a} )  
\chi^{\cB }_{j_a} \Bigr)   \Bigg]_{\sigma ,\chi , \bar\chi =0 } \; .  
\eeann 
In order to compute the derivatives with respect to the block Fermionic fields $\chi^{\cB}_j$ and $\bar \chi^{\cB}_j$ 
we note that such a derivative acts only on 
$ \prod_{a\in \cB} \Bigl(  \chi^{\cB}_{j_a}   \bar \chi^{ \cB  }_{j_a}  \Bigr)  $ and, furthermore, 
\[
  \frac{\partial}{\partial \bar \chi_j^{\cB} } 
  \prod_{a\in \cB} \Bigl(  \chi^{\cB}_{j_a}   \bar \chi^{ \cB  }_{j_a}  \Bigr)  
  = \Bigl( \sum_{a'\in \cB} \delta_{jj_{a'}}  \frac{\partial}{\partial \bar \chi_{j_{a'}}^{\cB} }  \Bigr)
  \prod_{a\in \cB} \Bigl(  \chi^{\cB}_{j_a}   \bar \chi^{ \cB  }_{j_a}  \Bigr)  
 \; ,  \;\;
 \frac{\partial}{\partial \chi_j^{\cB} } 
  \prod_{a\in \cB} \Bigl(  \chi^{\cB}_{j_a}   \bar \chi^{ \cB  }_{j_a}  \Bigr)  
  = \Bigl( \sum_{a'\in \cB} \delta_{jj_{a'}}  \frac{\partial}{\partial   \chi_{j_{a'}}^{\cB} }  \Bigr)
  \prod_{a\in \cB} \Bigl(  \chi^{\cB}_{j_a}   \bar \chi^{ \cB  }_{j_a}  \Bigr)  \; .
\] 
It follows that the Grassmann Gaussian integral is:
\beann
 && \Bigg[  e^{ 
   \sum_{\cB,\cB'} Y_{\cB\cB'}(w_{\ell_F})\sum_{a\in \cB, b\in \cB'} \delta_{j_aj_b}
     \frac{\partial}{\partial \bar \chi_{j_a}^{\cB} } \frac{\partial}{\partial \chi_{j_b}^{\cB'} } } \crcr 
  &&      \prod_{L_F \in \cL_F} 
   \Bigg(\sum_{a\in \cB(L_F),b\in \cB'(L_{F})}  \delta_{j_a j_b}\Big( \frac{\partial}{\partial \bar \chi_{j_a}^{\cB(L_F)} } 
   \frac{\partial}{\partial \chi_{j_b}^{\cB'(L_F)} } +  
   \frac{\partial}{\partial \bar \chi_{j_b}^{\cB'(L_F)} } \frac{\partial}{\partial \chi_{j_a}^{\cB(L_F)} } \Big)
   \Bigg) \crcr
&&
  \qquad \qquad\qquad  \prod_{\cB} \prod_{a\in \cB} \Bigl(  \chi^{\cB}_{j_a}   \bar \chi^{ \cB  }_{j_a}  \Bigr)  \Bigg]_{\chi^{\cB}_j, \bar\chi^{\cB}_j =0 } \; .
\eeann
The sums over $  a\in \cB(\ell_F) $ and  $ b\in \cB'(\ell_F)$ yield a sum over all the possible ways to hook the edge $L_F\in \cL_F$ 
to vertices in its end blocks. Each term represents a detailed Fermionic edge $\ell_F$ in the original graph (having the same $w_{\ell_F}= w_{L_F}$ parameter).
The sum over $\cL_F$ becomes therefore a sum over detailed Fermionic forests $\cF_F$ in the original graph (in which the Bosonic blocks are not contracted)
and we obtain a two-level jungle formula \cite{AR1} for the partition function:
\[
Z(\lambda, N) =  \sum_{n=0}^\infty \frac{1}{n!}  \sum_{\cJ} \;\sum_{j_1=j_{\min}}^{j_{\max }  } 
  \dots \sum_{j_n=j_{\min}}^{j_{\max} }
 \;  \int dw_\cJ  \;  \int d\nu_{ \cJ}  
\quad   \partial_\cJ   \Big[ \prod_{\cB} \prod_{a\in \cB}   \Bigl(    W_{j_a}   (  \sigma^a_{j_a} )  
\chi^{ \cB }_{j_a}   \bar \chi^{\cB}_{j_a}  \Bigr)  \Big] \; ,
\]
where
\begin{itemize}

\item the sum over $\cJ$ runs over all two level jungles, hence over all ordered pairs $\cJ = (\cF_B, \cF_F)$ of two (each possibly empty) 
disjoint forests on $V$, such that 
$\bar \cJ = \cF_B \cup \cF_F $ is still a forest on $V$. The forests $\cF_B$ and $\cF_F$ are the Bosonic and Fermionic components of $\cJ$.
The edges of $\cJ$ are partitioned into Bosonic edges $\ell_B$ and Fermionic edges $\ell_F$.
 
\item  $\int dw_\cJ$ means integration from 0 to 1 over parameters $w_\ell$, one for each edge $\ell \in \bar\cJ$.
$\int dw_\cJ  = \prod_{\ell\in \bar \cJ}  \int_0^1 dw_\ell  $.
There is no integration for the empty forest since by convention an empty product is 1. A generic integration point $w_\cJ$
is therefore made of $\vert \bar \cJ \vert$ parameters $w_\ell \in [0,1]$, one for each $\ell \in \bar \cJ$.

\item 
\[ \partial_\cJ  = \prod_{\genfrac{}{}{0pt}{}{\ell_B \in \cF_B}{\ell_B=(c,d)}} \Bigl(  \frac{\partial}{\partial \sigma^{c}_{j_{ c}} } 
 \frac{\partial}{\partial \sigma^{d}_{j_{d }  } } \Bigr)
\prod_{\genfrac{}{}{0pt}{}{\ell_F \in \cF_F}{\ell_F=(a,b) } } \delta_{j_{a } j_{b } } \Big(
   \frac{\partial}{\partial \bar \chi^{\cB(a)}_{j_{a}  } }\frac{\partial}{\partial \chi^{\cB(b)}_{j_{b }  } }+ 
    \frac{\partial}{\partial \bar \chi^{ \cB( b) }_{j_{b} } } \frac{\partial}{\partial \chi^{\cB(a) }_{j_{a}  } }
   \Big) \; ,
\]
where $ \cB(a)$ denotes the Bosonic blocks to which $a$ belongs. 

\item The measure $d\nu_{\cJ}$ has covariance $ X (w_{\ell_B}) \otimes \bbone_\cS $ on Bosonic variables and $ Y (w_{\ell_F}) \otimes \mathbb{I}_\cS  $  
on Fermionic variables, 
\[
 e^{\frac{1}{2} \sum_{a,b=1}^n X_{ab}(w_{\ell_B }) \frac{\partial}{\partial \sigma^a_{j_a} }\frac{\partial}{\partial \sigma^b_{j_b}} 
   +  \sum_{\cB,\cB'} Y_{\cB\cB'}(w_{\ell_F})\sum_{a\in \cB,  b\in \cB' } \delta_{j_aj_b}
   \frac{\partial}{\partial \bar \chi_{j_a}^{\cB} } \frac{\partial}{\partial \chi_{j_b}^{\cB'} } } \; .
\] 

\item  $X_{ab} (w_{\ell_B} )$  is the infimum of the $w_{\ell_B}$ parameters for all the Bosonic edges $\ell_B$
in the unique path $P^{\cF_B}_{a \to b}$ from $a$ to $b$ in $\cF_B$. The infimum is set to zero if such a path does not exists and 
to $1$ if $a=b$. 

\item  $Y_{\cB\cB'}(w_{\ell_F})$  is the infimum of the $w_{\ell_F}$ parameters for all the Fermionic
edges $\ell_F$ in any of the paths $P^{\cF_B \cup \cF_F}_{a\to b}$ from some vertex $a\in \cB$ to some vertex $b\in \cB'$. 
The infimum is set to $0$ if there are no such paths, and to $1$ if such paths exist but do not contain any Fermionic edges.

\end{itemize}

Remember that a main property of the forest formula is that the symmetric $n$ by $n$ matrix $X_{ab}(w_{\ell_B})$ 
is positive for any value of $w_\cJ$, hence the Gaussian measure $d\nu_{\cJ} $ is well-defined. The matrix $Y_{\cB\cB'}(w_{\ell_F})$
is also positive, with all elements between 0 and 1. Since the slice assignments, the fields, the measure and the integrand are now 
factorized over the connected components of $\bar \cJ$, the logarithm of $Z$ is easily computed as exactly the same sum but restricted 
to the two-levels spanning trees:
\bea \label{treerep}  
\log Z(\lambda, N) =  \sum_{n=1}^\infty \frac{1}{n!}  \sum_{\cJ \;{\rm tree}} \;\sum_{j_1=1}^{j_{\text{max}}} 
  \dots \sum_{j_n=1}^{j_{\text{max}} }
 \;  \int dw_\cJ  \;  \int d\nu_{ \cJ}  
\quad   \partial_\cJ   \Big[ \prod_{\cB} \prod_{a\in \cB}   \Bigl(    W_{j_a}   (  \sigma^a_{j_a} )  
 \chi^{ \cB }_{j_a} \bar \chi^{\cB}_{j_a} \Bigr)    \Big] \; , 
\eea
where the sum is the same but conditioned on $\bar \cJ = \cF_B \cup \cF_F$ being a \emph{spanning tree} on $V= [1, \cdots , n]$.
Our main results are
\begin{theorem} \label{thetheorem} Fix $j_{\min}\ge 3$ and  $M \ge 10^8$.
The series \eqref{treerep} is absolutely convergent for $\lambda\in [-1,1]$ uniformly in $j_{\max}$.
\end{theorem}
 
\begin{theorem} \label{thm:theorem2} Fix $j_{\min}\ge 3$ and  $M \ge 10^8$. The series \eqref{treerep} is absolutely convergent for 
$\lambda\in \mathbb{C}$, $\lambda = |\lambda|e^{\imath \gamma}$ in the domain $    |\lambda|^2 <   (\cos2\gamma) $ uniformly in $j_{\max}$. 
\end{theorem}

The restriction $j_{\min}\ge 3$ can be lifted easily: both  conditions $j_{\min}\ge 3$ and  $M \ge 10^8$ are not optimal and 
were chosen for the simplicity of the resulting domain in $\lambda$.

\section{Proof of Theorem \ref{thetheorem}}

In this section we prove our main theorems. We start with theorem \ref{thetheorem} and subsequently proceed to theorem \ref{thm:theorem2}. 
In eq. \eqref{treerep} the Bosonic and the Fermionic integrals 
decouple. Furthermore the Bosonic integral factors over the Bosonic blocks $\cB$,
\bea\label{eq:stage1}
  && \log Z(\lambda, N)  = \sum_{n=1}^\infty \frac{1}{n!}   
   \sum_{\cF_B, \cF_F }^{   \cF_B\cup \cF_F \text{ connected}}
   \; \sum_{j_1=j_{\min}}^{j_{\max} } \dots \sum_{j_n=j_{\min}}^{j_{\max} } \;
      \int_{0}^1 \prod_{\ell_B\in \cF_B } dw_{\ell_B} \prod_{\ell_F\in \cF_F } dw_{\ell_F}
  \crcr
&&  \qquad \qquad  \times \Bigg[ 
e^{\sum_{\cB,\cB'} Y_{\cB\cB'}(w_{\ell_F})\sum_{a\in \cB,  b\in \cB' } \delta_{j_aj_b}
   \frac{\partial}{\partial \bar \chi_{j_a}^{\cB} } \frac{\partial}{\partial \chi_{j_b}^{\cB'} } } \crcr
&& \qquad \qquad \qquad \qquad \qquad  \times \prod_{\genfrac{}{}{0pt}{}{\ell_F \in \cF_F}{\ell_F=(a,b) } } \delta_{j_{a } j_{b } } \Big(
   \frac{\partial}{\partial \bar \chi^{\cB(a)}_{j_{a}  } }\frac{\partial}{\partial \chi^{\cB(b)}_{j_{b }  } }+ 
    \frac{\partial}{\partial \bar \chi^{ \cB( b) }_{j_{b} } } \frac{\partial}{\partial \chi^{\cB(a) }_{j_{a}  } }
   \Big)   \prod_{\cB} \prod_{a\in \cB}   \Bigl( \chi^{ \cB }_{j_a} \bar \chi^{\cB}_{j_a} \Bigr) \Bigg]_{\chi \bar\chi =0 } \crcr
&&  \qquad \qquad  \times  \prod_{\cB} \Bigg[ 
e^{\frac{1}{2} \sum_{a,b\in \cB} X_{ab}(w_{\ell_B }) \frac{\partial}{\partial \sigma^a_{j_a} }\frac{\partial}{\partial \sigma^b_{j_b}} 
  }  
  \prod_{\genfrac{}{}{0pt}{}{\ell_B \in \cF_B \cap \cB }{\ell_B=(c,d)}} \Bigl(  \frac{\partial}{\partial \sigma^{c}_{j_{ c}} } 
 \frac{\partial}{\partial \sigma^{d}_{j_{d }  } } \Bigr)
    \prod_{a\in \cB}   W_{j_a}   (  \sigma^a_{j_a} )  \Bigg]_{\sigma  =0 } \; .
\eea
In this formula we have vertex replicas for the Bosonic variables and block Fermionic variables.
Both the Fermionic and the Bosonic variables of the vertex $a$ have the same scale $j_a$.
 
The proof of the theorem is divided into four parts. In subsection \ref{subsec:counting} we explain how the sum over two level trees is performed.
In subsection \ref{subsec:fermi} we evaluate the Grassmann Gaussian integral. In subsection \ref{subsec:boso} we deal with the Bosonic Gaussian integral
and finally in subsection \ref{subsec:final} we establish the convergence of the series in eq. \eqref{treerep}. 

\subsection{Counting two level trees}\label{subsec:counting}

Recall that the number of partitions of the set $\{1,\dots ,n \}$ into $m_1$ blocks of size $1$, $m_2$ blocks of size $2$,
and so on is 
\[
  \frac{n!}{\prod_{q\ge 1} m_q! (q!)^{m_q}  } \; ,
\]
and the number of trees over $q$ vertices labelled $1,\dots q$, with assigned coordination $d_i$ of the vertex $i$ is 
\[
\frac{(q-2)!}{(d_1-1)!\dots (d_q-1)!} \;, \qquad \sum_{i=1}^q d_i = 2q-2 \; .
\]
The sum over two level trees can be reorganized as follows:
\begin{itemize}
 \item we chose a partition $\cP$  of the set of vertices $\{1,\dots n\}$ into subsets $\cB$:
 \[
  \cP= \Big{\{} \cB \Big{|} \cB \subset  \{1,\dots n\} \Big{\}} \text{ such that } 
  \begin{cases}
\forall \cB, \cB' \in \cP \; , \qquad \cB \neq \cB' \Rightarrow \cB\cap \cB' =\emptyset \\   
  \cup_{\cB \in \cP } \cB = \{1,\dots n\}   
\end{cases} \; .
 \]
      A subset $\cB$ is a Bosonic block, and its cardinal is denoted $|\cB|$. We denote $|\cP|$ the number of blocks in the partition. 
 \item we chose a set of Fermionic edges $\cL_F$ forming a tree between the Bosonic blocks, with assigned coordination $D_1,\dots D_{|\cP|}$ at each block.
 \item we chose a way of hooking each Fermionic edge to a particular vertex in a block. Each edge has $|\cB|$ choices to hook at the block $\cB$.
       Each such choice leads to a detailed Fermionic forest $\cF_F$.
 \item we chose a Bosonic tree $\cT\in \cB$ in each Bosonic block.
\end{itemize}
As the contribution of a forest factors over the Bosonic blocks, the sum over two level trees becomes:
\[
\sum_{\cF_B,\cF_F}^{\cF_B \cup \cF_F \text{ connected}} A(\cF_F) \prod_{\cB} A(\cB) = \sum_{\cP} \sum_{\cF_F} A(\cF_F) 
\Bigl( \prod_{\cB} \sum_{\cT\in \cB} A(\cB) \Bigr) \; .
\]

We can for instance compute the number of two level trees over $n$ vertices. The contribution of each two level tree in the sum above is $1$ (which 
factors over the Bosonic blocks) hence the number of two level trees is:
\[
 \sum_{ \cP  }
  \sum_{D_{\cB}\ge 1, \forall \cB\in \cP}^{\sum_{\cB\in \cP} D_{\cB} - |\cP| = |\cP|-2}
   \frac{(|\cP|-2)!}{\prod_{\cB\in \cP} (D_{\cB}-1)!} \; \Bigl( \prod_{ \cB \in \cP} |\cB|^{D_B} \Bigr)  
  \quad  \Bigl( \prod_{ \cB \in \cP} |\cB|^{|\cB|-2} \Bigr)   \; ,
\] 
where the first factor counts the number of Fermionic trees with assigned coordinations, the second one takes into account that  
each Fermionic edge has $|\cB|$ choices to hook at the block $\cB$, and the last one counts the number of Bosonic trees inside each Bosonic block. 
The sums over $D_{\cB}$ can be computed and we get:
\bea\label{eq:counttwolevel}
 \sum_{\cP} \Bigl( \sum_{\cB \in \cP }  |\cB|\Bigr)^{|\cP|-2} \Bigl( \prod_{ \cB \in \cP} |\cB|^{|\cB|-1} \Bigr) 
 = \sum_{B_1,\dots B_q \dots \ge 0}^{\sum_{q\ge 1} qB_q =n}    \frac{n!}{\prod_{q\ge 1} B_q! (q!)^{B_q} }  n^{\sum_{q\ge 1} B_q-2}
   \prod_{q\ge 1}\Bigl(    q^{q-1}  \Bigr)^{B_q}
 \; ,
\eea 
where we grouped together the contributions of all the partitions having $B_1$ blocks of size $1$, $B_2$ blocks of size $2$, $B_q$ blocks of size $q$ and so on.
\begin{prop}\label{prop:counting}
 The number of two level trees over $n\ge 1$ vertices is bounded by $2^{2n} n^{n-2}$.
\end{prop}
\noindent{\bf Proof:} The proposition is trivial, as the number of such trees is exactly $2^{n-1}n^{n-2}$. It is however instructive to 
derive a bound directly starting from eq. \eqref{eq:counttwolevel}. We first use $\frac{q^{q-1}}{(q-1)!} \le e^q$ to obtain a bound:
\[
 \frac{n!}{n^2} \; e^n  \sum_{B_1,\dots B_q \dots \ge 0}^{\sum_{q\ge 1} qB_q =n}  
 \frac{1}{\prod_{q\ge 1} B_q! \; q^{B_q} }  n^{\sum_{q\ge 1} B_q} \; .
\]
The sum over $B_1,\dots B_q,\dots$ is nothing but the coefficient of $x^n$ in the Taylor expansion  
\[
\prod_{q\ge 1} \sum_{B_q\ge 0} \frac{1}{B_q!} \Bigl(n\frac{x^q}{q} \Bigr)^{B_q} = \prod_{q\ge 1} e^{n\frac{x^q}{q}} =  e^{n [ - \ln(1-x) ] }= \frac{1}{(1-x)^n}
= \sum_{p\ge 0} \binom{p+n-1}{p} x^p 
\;.
\]
The Stirling formula provides a tight bound on the factorial $\sqrt{2\pi n} \Bigl( \frac{n}{e} \Bigr)^n e^{\frac{1}{12n+1}} \le n! \le\sqrt{2\pi n} \Bigl( \frac{n}{e} \Bigr)^n e^{\frac{1}{12n}}  $,
hence the number of two level trees over $n$ vertices is bounded by 
\[
 \frac{1}{n^2}  e^n \frac{(2n-1)!}{(n-1)!}=\frac{1}{n } e^n \frac{(2n-1)!}{n!}
 \le \frac{1}{n}   \sqrt{2} \;  e^{\frac{1}{12}} \; \frac{(2n-1)^{2n-1}}{n^{n}} \le   \sqrt{2} \;  e^{\frac{1}{12}} \; \frac{ (2n)^{2n-1}}{n^{n+1}} <
  2^{2n } n^{n-2} \; .
\] 

\qed

Going back to our problem, the expression eq. \eqref{eq:stage1} is reorganized in terms of partitions $\cP$ of the set of vertices as: 
\bea\label{eq:stage2}
  && \log Z(\lambda, N)  = \sum_{n=1}^\infty \frac{1}{n!}   \sum_{\cP}
    \sum_{\cF_F} \;\; \sum_{ \{j_a\}} \;\; \int_0^1 \prod_{\ell_F\in \cF_F } dw_{\ell_F} \\
    &&  \qquad \times  \Bigg[ 
e^{\sum_{\cB,\cB'} Y_{\cB\cB'}(w_{\ell_F})\sum_{a\in \cB,  b\in \cB' } \delta_{j_aj_b}
   \frac{\partial}{\partial \bar \chi_{j_a}^{\cB} } \frac{\partial}{\partial \chi_{j_b}^{\cB'} } } \crcr
&& \qquad \qquad \qquad \qquad \qquad  \times \prod_{\genfrac{}{}{0pt}{}{\ell_F \in \cF_F}{\ell_F=(a,b) } } \delta_{j_{a } j_{b } } \Big(
   \frac{\partial}{\partial \bar \chi^{\cB(a)}_{j_{a}  } }\frac{\partial}{\partial \chi^{\cB(b)}_{j_{b }  } }+ 
    \frac{\partial}{\partial \bar \chi^{ \cB( b) }_{j_{b} } } \frac{\partial}{\partial \chi^{\cB(a) }_{j_{a}  } }
   \Big)   \prod_{\cB} \prod_{a\in \cB}   \Bigl( \chi^{ \cB }_{j_a} \bar \chi^{\cB}_{j_a} \Bigr) \Bigg]_{\chi \bar\chi =0 } \crcr    
&& \qquad \times \prod_{\cB} \Bigg[ \sum_{\cT\in \cB}   
      \int_{0}^1 \prod_{\ell_B\in \cT } dw_{\ell_B}   e^{\frac{1}{2} \sum_{a,b\in \cB} X_{ab}(w_{\ell_B }) \frac{\partial}{\partial \sigma^a_{j_a} }\frac{\partial}{\partial \sigma^b_{j_b}} 
  }  
  \prod_{\genfrac{}{}{0pt}{}{\ell_B \in \cT }{\ell_B=(c,d)}} \Bigl(  \frac{\partial}{\partial \sigma^{c}_{j_{ c}} } 
 \frac{\partial}{\partial \sigma^{d}_{j_{d }  } } \Bigr)
    \prod_{a\in \cB}   W_{j_a}   (  \sigma^a_{j_a} )  \Bigg]_{\sigma  =0 } \; , \nonumber
\eea
 where $\sum_{\{j_a\}}$ signifies a sum over the slice indices of all the vertices, $j_a$, from $j_{\min}$ to $j_{\max}$.
 
\subsection{The Grassmann Gaussian integral}\label{subsec:fermi}

In order to evaluate further the Grassmann Gaussian integral 
we first note that any polynomial depending on pairs of Grassmann numbers $\chi_{\alpha}, \bar \chi_{\alpha}$ can be written as
$
P(\chi, \bar\chi) = \sum_{t\ge 0}  P(t) \; \chi_1 \dots \chi_{t}  \bar \chi_t   \dots \bar \chi_1 \;,
$
therefore:
\beann
 && P(\partial_{ \bar \chi_{\alpha} } ,\partial_{\chi_\alpha }) Q(\chi_{\alpha}, \bar \chi_{\alpha}) = \sum_{t,s\ge 0} P(t) Q(s) 
  \Bigl( \partial_{\bar \chi_1} \dots \partial_{\bar \chi_t} \partial_{\chi_t} \dots \partial_{\chi_1} \Bigr)
  \chi_1 \dots \chi_{s}  \bar \chi_t   \dots \bar \chi_s \crcr
  && \qquad = \sum_{t,s} P(t)\delta_{ts} Q(s) = Q(\partial_{ \bar \chi_{\alpha} } ,\partial_{\chi_\alpha }) P(\chi_{\alpha}, \bar \chi_{\alpha}) \; .
\eeann
The Gaussian Grassmann integral can thus be rewritten as:
\[
 \Bigg[ \prod_{\cB}\prod_{a\in \cB} \Bigl( \frac{\partial}{\partial \bar \chi^{\cB }_{j_a} }  \frac{ \partial}{\partial \chi^{\cB }_{j_a}} \Bigr)  
  e^{  \sum_{ \cB\cB' } Y_{\cB \cB' }(w_{\ell_F}) \sum_{a\in \cB,b\in \cB'} \delta_{j_aj_b}   \chi^{\cB}_{j_a}  \bar \chi^{\cB'}_{j_b} }
    \prod_{\genfrac{}{}{0pt}{}{\ell_F \in \cF_F}{\ell_F=(a,b) } } \delta_{j_{a } j_{b } } \Big(
    \chi^{\cB(a)}_{j_{a}  } \bar  \chi^{\cB(b)}_{j_{b } }  + 
     \chi^{ \cB( b) }_{j_{b} }  \bar  \chi^{\cB(a) }_{j_{a} } \Big)   
  \Bigg]_{\chi, \bar\chi = 0} \; .
\]
There are exactly $2n$ distinct Fermionic fields $\chi^{\cB}_{j_a}, \bar \chi^{\cB}_{j_a}$, therefore any polynomial of global degree
greater than $2n$ is automatically zero and the restriction to $\chi^{\cB}_{j_a}=\bar \chi^{\cB}_{j_a}=0$ is automatically fulfilled. 
Denoting ${\bf Y}_{ab} = Y_{\cB(a) \cB(b)} (w_{\ell_F}) \delta_{j_a j_b} $, 
and taking into account that $  Y_{\cB\cB'}(w_{\ell_F}) $ is symmetric, the above Gaussian integral takes the more familiar form:
\[  \int \prod_{\cB} \prod_{a\in \cB}  ( d  \bar \chi^{\cB}_{j_a}  d  \chi^{\cB}_{j_a}   )
  \; \; e^{  - \sum_{a,b=1}^n    \bar \chi^{\cB(a)}_{j_a} {\bf Y}_{ab}  \chi^{\cB(b)}_{j_b}  }
\prod_{\genfrac{}{}{0pt}{}{\ell_F \in \cF_F}{\ell_F=(a,b) } } \delta_{j_{a } j_{b } } \Big(
    \chi^{\cB(a)}_{j_{a}  } \bar  \chi^{\cB(b)}_{j_{b } }  + 
     \chi^{ \cB( b) }_{j_{b} }  \bar  \chi^{\cB(a) }_{j_{a} } \Big) \;.  
\] 

The important observation is that this integral obeys a hard core constraint inside each block: if two vertices $a$ and $b$ belong to the same 
Bosonic block $\cB$ and have the same scale $j_a =j_b$, then the integral is zero (as we integrate twice with respect to the same Grassmann
variable $ \chi^{\cB}_{j_a} = \chi^{\cB}_{j_b}$). To emphasize this constraint, we multiply the integral 
by $ \prod_{\genfrac{}{}{0pt}{}{a,b\in \cB}{a\neq b}} (1-\delta_{j_aj_b})$,
which is zero unless the slices indices $j_a$ of the vertices of the block $\cB$ are all different. 
We furthermore denote $k$ the number of edges in the Fermionic forest $\cF_F$ and for any matrix $M$ we denote:
 \[
   M^{ \hat b_1 \dots   \hat b_k}_{  \hat a_1 \dots   \hat a_k} = 
   \int (\prod_i d\bar \psi_i d\psi_i) \; e^{- \sum_{i,j}\bar \psi_i M_{ij} \psi_j} \prod_{i=1}^k \psi_{a_i} \bar  \psi_{b_i} \; ,
  \]
which is (up to a sign) the minor of $M$ with the lines $b_1\dots b_k$ and the columns $a_1\dots a_k$ deleted. The Grassmann Gaussian integral evaluates to: 
\bea\label{eq:grassmaint}
 \Bigl( \prod_{\cB} \prod_{\genfrac{}{}{0pt}{}{a,b\in \cB}{a\neq b}} (1-\delta_{j_aj_b}) \Bigr)
 \Bigl( \prod_{\genfrac{}{}{0pt}{}{\ell_F \in \cF_F}{\ell_F=(a,b) } } \delta_{j_{a } j_{b } } \Bigr)
 \Bigl( {\bf Y }^{\hat b_1 \dots \hat b_k}_{\hat a_1 \dots \hat a_k}  + 
 {\bf Y }^{\hat a_1 \dots \hat b_k}_{\hat b_1 \dots \hat a_k}+\dots + {\bf Y }_{\hat b_1 \dots \hat b_k}^{\hat a_1 \dots \hat a_k}   \Bigr) \; ,
\eea 
where the sum runs over the $2^k$ ways to exchange an $a_i$ and a $b_i$. 
\begin{lemma}\label{lem:ferm}
 For any $a_1,\dots a_k$ and $b_1,\dots b_k$,
 \[
   \Big{|}  {\bf Y }^{\hat b_1 \dots \hat b_k}_{\hat a_1 \dots \hat a_k} \Big{|}\le 1 \; .
 \]
\end{lemma}

The proof of this statement is the combined result of the following two propositions.

\begin{prop}
 The matrix ${\bf Y}$ has real (positive) entries and unit diagonal entries ${\bf Y}_{aa}=1$. 
 Furthermore, ${\bf Y}$ is a positive matrix.
\end{prop}
{\bf Proof:} The first statement is trivial. For the second statement,
consider a fixed set of $w_{\ell_F}$ parameters, ordered in decreasing order 
$  w_1 \ge w_2 \ge \dots w_{k}\ge 0$. By convention we define $w_0 :=1$ and $w_{k+1}:=0$. 
The first $i$ edges $w_1, \dots w_i$ yield a set of Fermionic trees at level $i$, $F^{(i)}$,
whose vertices are the Bosonic blocks. Two Bosonic blocks $\cB$ and $\cB'$ belong to the same Fermionic 
tree $\cB,\cB' \in f\in F^{(i)}$ iff any two vertices $a\in \cB$ and $b\in \cB'$ are connected by a path
in $\cF_B \cup \cF_F$ whose Fermionic edges are a subset of $w_1,\dots w_i$. 

Let $ ( x_{j_1}^{\cB(1)}, \dots , x_{j_n}^{\cB(n)} ) $ be some vector in $\mathbb{R}^n$. 
The quadratic form $ \sum_{a,b} x^{\cB(a)}_{j_a} {\bf Y}_{ab} x^{\cB(b)}_{j_b}$
writes in the manifestly positive manner \cite{AR1}
\bea\label{eq:telescope}
&& \sum_{a,b} x^{\cB(a)}_{j_a} {\bf Y}_{ab} x^{\cB(b)}_{j_b} = 
   \sum_{\cB,\cB'} \sum_{a\in \cB, b\in \cB'}  x^{\cB}_{j_a}  \Bigl( Y_{\cB \cB'}(w_{\ell_F}) \delta_{j_aj_b} \Bigr) x^{\cB'}_{j_b} \crcr
&&  = \sum_{i=0}^{k} (w_i - w_{i+1}) \sum_{f\in F^{(i)}} \Bigg[ \sum_{\cB,\cB'\in f}
   \Bigl( \sum_{a\in \cB,b\in \cB'} x^{\cB}_{j_a} \delta_{j_aj_b} x^{\cB'}_{j_b} \Bigr)  \Bigg] \crcr
&&    = \sum_{i=0}^{k} (w_i - w_{i+1}) \sum_{f\in F^{(i)}} \sum_{j=j_{\min}}^{j_{\max}}
   \Bigl(\sum_{\cB\in f} \sum_{a\in \cB} x^{\cB}_{j_a} \delta_{j_aj} \Bigr)  
   \Bigl( \sum_{ \cB'\in f}  \sum_{ b\in \cB'}  \delta_{j j_b} x^{\cB'}_{j_b}   \Bigr) 
   \; .
\eea 

Indeed, consider two blocks $\cB$ and $\cB'$. For all $i = 1,\dots n$, $ \cB \in f^{\cB,i}\in F^{(i)}$ and $ \cB' \in f^{\cB',i}\in F^{(i)}$. 
As $\cB$ and $\cB'$ are connected by $\cF_F$, there exists a $q$
such that $ f^{\cB,i}\neq f^{\cB',i},\; \forall i <q $ and $ f^{\cB,i} = f^{\cB',i},\; \forall i\ge q $. 
The coefficient of $ \sum_{a\in \cB, b\in \cB'} x^{\cB}_{j_a}  \delta_{j_aj_b} x^{\cB'}_{j_b} $ in the second line of eq.
\eqref{eq:telescope} is $w_q$, which is also the smallest $w$ of a Fermionic edge in any path in $\cF_B\cup \cF_F$ connecting a 
vertex in $\cB$ with a vertex in $\cB'$, and therefore it equals $Y_{\cB\cB'}(w_{\ell_F})$.

\qed

\begin{prop}
 Let $M$ be a real, symmetric, positive matrix. Then 
  \bea\label{eq:part1}
   \Big( M^{\hat b_1\dots \hat b_k}_{\hat a_1\dots \hat a_k} \Big)^2 \le 
   M^{\hat a_1\dots \hat a_k}_{\hat a_1\dots \hat a_k}  \; \;  M^{\hat b_1\dots \hat b_k}_{\hat b_1\dots \hat b_k}  \; .
  \eea 
  If, furthermore, $M$ has unit diagonal entries, then 
  \bea\label{eq:part2}
    M^{\hat a_1\dots \hat a_k}_{\hat a_1\dots \hat a_k}  \le 1 \; .
  \eea 
\end{prop}
{\bf Proof:} The diagonal minors $ M^{\hat a_1\dots \hat a_k}_{\hat a_1\dots \hat a_k}  $ are real and positive (each minor is the determinant
of the restriction of $M$ to some subspace, which is still a positive matrix). As $M$ is symmetric, 
$ M^{\hat b_1\dots \hat b_k}_{\hat a_1\dots \hat a_k} = M_{\hat b_1\dots \hat b_k}^{\hat a_1\dots \hat a_k} $.
We consider the Grassmann integral:
\[
I :=  \int \prod_i (d\bar \psi_i d\psi_i) \; e^{- \sum_{i,j}\bar \psi_i M_{ij} \psi_j} 
 \Bigl( \psi_{a_1} \dots \psi_{a_k} + \lambda \psi_{b_1} \dots \psi_{b_k} \Bigr) 
 \Bigl(  \bar \psi_{a_k} \dots \bar \psi_{a_1} + \lambda \bar \psi_{b_k} \dots \bar \psi_{b_1} \Bigr) \; ,
\]
for any real $\lambda$. On the one hand this integral evaluates explicitly to:
\bea \label{eq:cauch1}
  M^{\hat a_1\dots \hat a_k}_{\hat a_1\dots \hat a_k} + 2 \lambda \;   M^{\hat b_1\dots \hat b_k}_{\hat a_1\dots \hat a_k} + 
  \lambda^2 \; M^{\hat b_1\dots \hat b_k}_{\hat b_1\dots \hat b_k}  \; .
\eea 
On the other, it is  positive for all real $\lambda$. To see this, we first rewrite it as
\beann
&&  \int \prod_i (d\bar \psi_i d\psi_i) \; e^{- \sum_{i,j}\bar \psi_i M_{ij} \psi_j} \crcr
&& \qquad \qquad \times  \frac{1}{k!} \Bigg( \sum_{\sigma\in \mathfrak{S}_k } \epsilon(\sigma)
  \Bigl( \prod_{l=1}^k\psi_{a_{\sigma(l)} } + \lambda \prod_{l=1}^k \psi_{b_{\sigma(l)} }
  \Bigr) \Bigg) 
  \frac{1}{k!} \Bigg( \sum_{\tau\in \mathfrak{S}_k } \epsilon(\tau )
  \Bigl( \prod_{l=k}^1 \bar \psi_{a_{\tau(l)} } + \lambda \prod_{l=k}^1 \bar \psi_{b_{\tau(l)} }
  \Bigr) \Bigg) \; . 
\eeann 
Being symmetric, the matrix $M$ can be diagonalized by an orthogonal transformation $M = O D O^T$, and all its eigenvalues $m_i$ are positive. 
By the change of variables of unit Jacobian $\chi_i = \sum_j O_{ji} \psi_j, \bar \chi_j  = \sum_i \bar \psi_i O_{ij}$, we get:
\beann
&& I= \frac{1}{k!k!} \sum_{\sigma,\tau\in \mathfrak{S}_k} \epsilon(\sigma) \epsilon(\tau)
 \sum_{  \genfrac{}{}{0pt}{}{i_1,\dots i_k}{ p_1,\dots p_k }} \Bigl(  \prod_{l=1}^k  O_{a_{\sigma(l)}i_l} + \lambda \prod_{l=1}^kO_{b_{\sigma(l)}i_l} \Bigr)
  \Bigl(  \prod_{l=k}^1  O_{a_{\tau(l)}p_l} + \lambda \prod_{l=k}^1 O_{b_{\tau(l)}p_l} \Bigr)   \crcr
&& \qquad \qquad \qquad \times \int \prod_i (d\bar \chi_i d\chi_i) e^{-\bar \chi_i m_i \chi_i}  \chi_{i_1}\dots \chi_{i_k}   
\bar \chi_{p_k}\dots \bar \chi_{p_1} \; .  
\eeann
The integrals over $\bar\chi, \chi$ yield:
\[
 \int \prod_i (d\bar \chi_i d\chi_i) e^{-\bar \chi_i m_i \chi_i}    \;\; \chi_{i_1}\dots \chi_{i_k} \bar \chi_{p_k}\dots \bar \chi_{p_1}
 = \sum_{\pi \in \mathfrak{S}_k } \epsilon(\pi) \prod_{l=1}^k \delta_{i_{l}  p_{ \pi(l) }} \prod_{\genfrac{}{}{0pt}{}{i=1}{ i\neq i_1\dots i_k}}^n m_i \; ,
\]
and finally:
\beann
&& I = \sum_{   i_1,\dots i_k } \Bigl( \prod_{\genfrac{}{}{0pt}{}{i=1}{ i\neq i_1\dots i_k}}^n m_i \Bigr) \frac{1}{k!k!} \crcr
&& \qquad \qquad   \sum_{\sigma,\tau,\pi  \in \mathfrak{S}_k  } \epsilon(\sigma)\epsilon(\tau) \epsilon(\pi)
  \Bigl(  \prod_{l=1}^k  O_{a_{\sigma(l)}i_l} + \lambda \prod_{l=1}^kO_{b_{\sigma(l)}i_l} \Bigr)
  \Bigl(  \prod_{l=k}^1  O_{a_{\tau\pi(l)}p_l} + \lambda \prod_{l=k}^1 O_{b_{\tau\pi(l)}p_l} \Bigr) 
    \crcr
&& \quad =  \sum_{   i_1,\dots i_k } \Bigl( \prod_{\genfrac{}{}{0pt}{}{i=1}{ i\neq i_1\dots i_k}}^n m_i \Bigr) \frac{1}{k! }
  \Bigl[ \sum_{\sigma} \epsilon (\sigma) \Bigl( \prod_{l=1}^k O_{a_{\sigma(l) } i_l }  + \lambda  O_{b_{\sigma(l) } i_l }  \Bigr) \Bigr]^2 \ge 0 \; . 
\eeann

It follows that the discriminant of eq. \eqref{eq:cauch1}, seen as a polynomial in $\lambda$, must be negative hence the first part of the lemma, 
eq. \eqref{eq:part1} holds. To establish eq. \eqref{eq:part2} we note that any positive real matrix $R$ admits a square root $R=Z^2$, and by the Hadamard inequality,
\[
 \det R = (\det Z)^2 \le \Bigl[ \prod_{i=1}^n (\sum_k z_{ik}z_{ki} )\Bigr]  =  \Bigl[ \prod_{i=1}^n R_{ii}\Bigr]^2 \; .
\] 
In the case at hand in eq. \eqref{eq:part2}, the diagonal $R_{ii}$ terms are all equal to 1.

\qed

Substituting eq. \eqref{eq:grassmaint} in eq. \eqref{eq:stage2} the logarithm of the partition function becomes:
\bea\label{eq:stage3}
  && \log Z(\lambda, N)  = \sum_{n=1}^\infty \frac{1}{n!}   \sum_{\cP}
    \sum_{\cF_F} \;\; \sum_{ \{j_a\} } \;\;    \Bigl( \prod_{\genfrac{}{}{0pt}{}{\ell_F \in \cF_F}{\ell_F=(a,b) } } \delta_{j_{a } j_{b } } \Bigr)  \\
  && \qquad \times 
   \int_0^1 \prod_{\ell_F\in \cF_F } dw_{\ell_F} 
        \prod_{\genfrac{}{}{0pt}{}{\ell_F \in \cF_F}{\ell_F=(a,b) } }  
 \Bigl( {\bf Y }^{\hat b_1 \dots \hat b_k}_{\hat a_1 \dots \hat a_k}  + 
 {\bf Y }^{\hat a_1 \dots \hat b_k}_{\hat b_1 \dots \hat a_k}+\dots + {\bf Y }_{\hat b_1 \dots \hat b_k}^{\hat a_1 \dots \hat a_k}   \Bigr) \crcr
&& \qquad \times \prod_{\cB} \Bigg[ \sum_{\cT\in \cB}    
    \prod_{\genfrac{}{}{0pt}{}{a,b\in \cB}{a\neq b}} (1-\delta_{j_aj_b})  \int_{0}^1 \prod_{\ell_B\in \cT } dw_{\ell_B} \crcr
&& \qquad \qquad \times    e^{\frac{1}{2} \sum_{a,b\in \cB} X_{ab}(w_{\ell_B }) \frac{\partial}{\partial \sigma^a_{j_a} }\frac{\partial}{\partial \sigma^b_{j_b}} 
  }  
  \prod_{\genfrac{}{}{0pt}{}{\ell_B \in \cT }{\ell_B=(c,d)}} \Bigl(  \frac{\partial}{\partial \sigma^{c}_{j_{ c}} } 
 \frac{\partial}{\partial \sigma^{d}_{j_{d }  } } \Bigr)
    \prod_{a\in \cB}   W_{j_a}   (  \sigma^a_{j_a} )  \Bigg]_{\sigma  =0 } \; . \nonumber
\eea

\subsection{The Bosonic Gaussian integral}\label{subsec:boso}
 
The Gaussian integral in a Bosonic block $\cB$ corresponding to a Bosonic tree $\cT$ writes as:
\[
  \Bigg[ 
e^{\frac{1}{2} \sum_{a,b\in \cB} X_{ab}(w_{\ell_B }) \frac{\partial}{\partial \sigma^a_{j_a} }\frac{\partial}{\partial \sigma^b_{j_b}} 
  }  
   \; \prod_{a\in \cT}  \Bigl( \frac{\partial }{    \partial  \sigma^a_{j_a}   }  \Bigr)^{ d_a }
  \;      W_{j_a}   (  \sigma^a_{j_a} )  \Bigg]_{\sigma  =0 } \; ,
\]
where $d_a$ is the coordination of the vertex $a$ in the Bosonic tree $\cT$. In order to evaluate the derivatives of the vertex kernels $W_j$ we will 
make use of the Fa\`a di Bruno formula (which is easily established by induction):
\[
 \partial^{(q)}_{x} f\bigl( g(x) \bigr) =   \sum_{\pi } f^{|\pi|}\bigl( g(x) \bigr) \prod_{B\in \pi} g^{|B|} (x)  \; ,
\]
where $\pi$ runs over the partitions of the set $\{1,\dots q\}$ and $B$ runs through the blocks of the partition $\pi$. Recalling that:
\[
  W_j= e^{-V_j}-1 \; , \qquad V_j (\sigma) = \sum_{p\in I_j}\bigl( \log(1- \imath \lambda \frac{\sigma}{p})  + \imath \lambda \frac{\sigma}{p} \bigr)  \; ,
\]
we obtain:
\beann
&& \partial_{\sigma}(- V_j ) =  \sum_{p\in I_j} \Bigl( \frac{\imath \lambda \frac{1}{p} }{ 1- \imath  \lambda \frac{\sigma}{p} } - \imath \lambda \frac{1}{p}\Bigr) 
  = \sum_{p\in I_j} \frac{ (-  \lambda^2 \sigma )}{p( p -  \imath \lambda \sigma )} \; ,\crcr
&& \partial^{(k)}_{\sigma}(- V_j ) = \sum_{p\in I_j}  
  \frac{ (k-1)! \Bigl( \imath\lambda\frac{1}{p} \Bigr)^k }{ \Bigl( 1- \imath \lambda\frac{\sigma}{p} \Bigr)^k} = 
   (k-1)! \sum_{p\in I_j}    \frac{\imath^k \lambda^k}{ (   p - \imath \lambda \sigma  )^k    }  \; ,  {\rm for}  \;\; k \ge 2\crcr
&&  \partial^{(q)}_{\sigma} W_j= e^{-V_j} \; 
 \sum_{m_1,m_2, \dots \ge 0}^{\sum_{k\ge 1} km_k = q} \frac{q!}{\prod_{k\ge 1}   m_k! (k!)^{m_k}  } \; 
  \prod_{k\ge 1} \; \bigl[ (-V_j)^{(k)} \bigr]^{m_k} \; ,
\eeann
and the Bosonic Gaussian integral becomes:
\bea\label{eq:bosogauss}
&& \Bigg[ 
 e^{\frac{1}{2} \sum_{a,b\in \cB} X_{ab}(w_{\ell_B }) \frac{\partial}{\partial \sigma^a_{j_a} }\frac{\partial}{\partial \sigma^b_{j_b}} 
  }  \prod_{a\in \cB}  e^{-V_{j_a} (\sigma_a) }  \\
&& \qquad    \prod_{a\in \cB} \;\;
   \sum_{m^{(a)}_1,m^{(a)}_2, \dots \ge 0}^{\sum_{k\ge 1} km^{(a)}_k = d_a} \frac{d_a!}{\prod_{k\ge 1}   m^{(a)}_k! (k!)^{m^{(a)}_k}  }
    \bigl[\sum_{p\in I_{j_a}} \frac{ (-  \lambda^2 \sigma^a_{j_a} )}{p( p -  \imath \lambda \sigma^a_{j_a} )}  \bigr]^{m^{(a)}_1} 
  \prod_{k\ge 2} \bigl[  \sum_{p\in I_{j_a}}    \frac{  (k-1)! \imath^k \lambda^k}{ (   p - \imath \lambda \sigma^a_{j_a}   )^k    }   \bigr]^{m^{(a)}_k} 
\Bigg]_{\sigma  =0 } \; . \nonumber
\eea 

\begin{lemma}\label{lem:boso}
 For $\lambda$ real, $|\lambda|<1$ the Bosonic Gaussian integral in eq. \eqref{eq:bosogauss} 
 is bounded by
 \[
  \sqrt{(4|\cB|-4)!!} \; \; \Bigl(  \prod_{a\in \cB}  d_a! \; |\lambda|^{d_a} \frac{1}{M^{  j_a-2 } }  \Bigr) \; , 
 \]
 where $|\cB|$ denotes the number of vertices of the block $\cB$.
\end{lemma}
\noindent{\bf Proof:} We rewrite eq. \eqref{eq:bosogauss} as
\beann
 && \sum_{\genfrac{}{}{0pt}{}{\forall a \in \cB} 
   {m^{(a)}_1,m^{(a)}_2, \dots \ge 0 }}^{\sum_{k\ge 1} km^{(a)}_k = d_a}
    \Bigl( \prod_{a\in \cB } \frac{d_a!}{\prod_{k\ge 1}   m^{(a)}_k!  \; k  ^{m^{(a)}_k}  }
    (\imath \lambda)^{m^{(a)}_1 + d_a  } \Bigr) \Bigg[ 
 e^{\frac{1}{2} \sum_{a,b\in \cB} X_{ab}(w_{\ell_B }) \frac{\partial}{\partial \sigma^a_{j_a} }\frac{\partial}{\partial \sigma^b_{j_b}} 
  } \\
 && \qquad \qquad \times \prod_{a\in \cB}  e^{-V_{j_a} (\sigma_a) }    (\sigma^a_{j_a} )^{m^{(a)}_1} 
  \bigl[ \sum_{p\in I_{j_a}} \frac{ 1}{p( p -  \imath \lambda \sigma^a_{j_a} )}  \bigr]^{m^{(a)}_1} 
  \prod_{k\ge 2} \bigl[  \sum_{p\in I_{j_a}}    \frac{ 1}{ (   p - \imath \lambda \sigma^a_{j_a}   )^k    }   \bigr]^{m^{(a)}_k}   \Bigg]_{\sigma  =0 } \; .
\eeann
Recalling that the Gaussian measure is positive, we use twice the Cauchy-Schwarz inequality to bound
\beann 
&& \Bigl{|}\int d\nu  \; \Bigl( \prod_a  e^{-V_{j_a} (\sigma_a) } \Bigr) \Bigl( \prod_a (\sigma^a_{j_a} )^{m^{(a)}_1} \Bigr) \Bigl( \prod_a F_a( \sigma^a_{j_a} ) \Bigr)
 \Bigr{|} \crcr
 && \qquad \le \Bigl{(} \int d\nu  \bigl{|}\prod_a (\sigma^a_{j_a} )^{m^{(a)}_1}  \bigr{|}^2 \Bigl{)}^{1/2} \; 
 \Bigl{(} \int d\nu   \bigl{|}\prod_a  e^{-V_{j_a} (\sigma_a) } \bigr{|}^4  \Bigr{)}^{1/4} \; \Bigl{(} \int d\nu  \bigl{|}\prod_a F_a( \sigma^a_{j_a} ) \bigr{|}^4 \Bigr{)}^{1/4} .
\eeann
Each term is bounded as follows. For the first term we evaluate the Gaussian integral and, as the normalized Gaussian measure has covariance smaller than $1$, 
each contraction is bounded by $1$ hence:
\[
 \int d\nu \prod_a  (\sigma^a_{j_a} )^{2m^{(a)}_1}  \le \Bigl(2 \sum_{a} m^{(a)}_1 \Bigr)!! \le   \Bigl(2 \sum_{a} d_a \Bigr)!!
  \le (4 |\cB| -4)!!
 \; ,
\]
where we have used $m_1^{(a)}\le d_a$ and $\sum_{a}d_a = 2 |\cB| -2$.
The second term is the simplest as $| e^{-V_{j_a} (\sigma_a) }|\le 1$ for $\lambda \in \mathbb{R}$. For the last term, we use 
\[
  \sum_{p\in I_j} \frac{1}{p^t} = \sum_{ p=M^{j-1} }^{M^j-1} \frac{1}{p^t} \le \frac{M^j}{M^{t(j-1)}} = \frac{M}{M^{(t-1)(j-1) } } \; ,
\]
hence a very rough bound is 
\beann
&& \bigl{|} F_a( \sigma^a_{j_a} ) \bigr{|} = \Bigl{|} \bigl[ \sum_{p\in I_{j_a}} \frac{ 1}{p( p -  \imath \lambda \sigma^a_{j_a} )}  \bigr]^{m^{(a)}_1} 
  \prod_{k\ge 2} \bigl[  \sum_{p\in I_{j_a}}    \frac{ 1}{ (   p - \imath \lambda \sigma  )^k    }   \bigr]^{m^{(a)}_k}    \Bigr{|} \crcr
&& \qquad \qquad \le \bigl[ \sum_{p\in I_{j_a}} \frac{ 1}{p^2 }  \bigr]^{m^{(a)}_1} 
  \prod_{k\ge 2} \bigl[  \sum_{p\in I_{j_a}}    \frac{ 1}{ p^k }   \bigr]^{m^{(a)}_k} 
  \le \frac{1}{M^{j_a-2}} \; ,
\eeann
as in the worst case $m_1^{(a)}=1$ and all the others are zero. Collecting all these bounds, we obtain
\beann
  \sum_{\genfrac{}{}{0pt}{}{\forall a \in \cB} 
   {m^{(a)}_1,m^{(a)}_2, \dots \ge 0 }}^{\sum_{k\ge 1} km^{(a)}_k = d_a} \sqrt{(4 |\cB| -4)!!} \prod_{a\in \cB }  
   \Bigg[ d_a! \; |\lambda|^{d_a}   
     \frac{ |  \lambda|^{m^{(a)}_1   }  }{\prod_{k\ge 1}   m^{(a)}_k!  \; k  ^{m^{(a)}_k}  }
   \frac{1}{M^{j_a-2}}
    \Bigg] \; . 
\eeann
Considering $|\lambda|\le 1$, hence $ |  \lambda|^{m^{(a)}_1   } \le 1$, the lemma follows taking into account that
\[
 \sum_{m_1,\dots m_k\dots \ge 0}^{\sum km_k =d} \frac{ 1}{\prod_k (m_k)! k^{m_k } } \; ,
\] 
is the coefficient of $x^d$ in the Taylor expansion 
\[
 \prod_{k\ge 1} e^{\frac{x^k}{k}} = e^{-[\ln (1-x)]} =\frac{1}{1-x} = \sum_{p\ge 0} x^p \; .
\]

\qed

\subsection{The final bound}\label{subsec:final}

Collecting eq. \eqref{eq:stage3}, lemma \ref{lem:ferm} and lemma \ref{lem:boso}, we get  the bound: 
\beann
  && \Big{|}\log Z(\lambda, N) \Big{|}  \le  \sum_{n=1}^\infty \frac{1}{n!}   \sum_{\cP}
    \sum_{\cF_F} \;\; \sum_{ \{j_a\} } \;\;    \Bigl( 2^{E(\cF_F)} \prod_{\genfrac{}{}{0pt}{}{\ell_F \in \cF_F}{\ell_F=(a,b) } } 
    \delta_{j_{a } j_{b } } \Bigr) 
    \crcr 
&& \qquad \qquad \times \prod_{\cB} \Bigg[ \sum_{\cT\in \cB}    
    \prod_{\genfrac{}{}{0pt}{}{a,b\in \cB}{a\neq b}} (1-\delta_{j_aj_b}) 
    \sqrt{(4 |\cB| -4)!!} \prod_{a\in \cB} \Bigl( d_a! \; |\lambda|^{d_a} \frac{1}{M^{ j_a-2 } } \Bigr) \Bigg] \; \;,
\eeann
where $ E(\cF_F)$ denotes the number of edges in the Fermionic forest. 

\begin{lemma}\label{lem:lastbound}
The logarithm of the partition function is bounded by 
\bea\label{eq:lastbound}
 \Big{|}\log Z(\lambda, N) \Big{|}  \le \sum_{B_1,\dots B_q,\dots \ge 0} \frac{ \bigl(\sum B_q \bigr)! }{\prod_q B_q! } \prod_{q\ge 1} 
 \Bigl[ |\lambda|^{2q-2} \; 3^{3q}  \;  q^q \; \frac{1}{M^{\frac{q^2}{4}}}     \Bigr]^{B_q}  \; ,
\eea 
 for $j_{\min}\ge 3$ and $M>4$.
\end{lemma}
\noindent{\bf Proof:} We first drop the constraint that the slice indices are conserved along the Fermionic edges. We can associate the sum over the slice index $j_a$
to the Bosonic block $\cB(a)$. The contributions of the trees $\cT$ in the Bosonic block can be summed together to yield the block contribution:
\[
 \sum_{ \{j_a\}, a\in \cB } \;  |\lambda|^{2|\cB|-2} \sqrt{ (4 |\cB| -4)!! } \; \; \Bigl( \prod_{\genfrac{}{}{0pt}{}{a,b\in \cB}{a\neq b}} (1-\delta_{j_aj_b}) \Bigr)
   \Bigl( \prod_{a\in \cB} \frac{ 1}   { M^{ j_a-2 }   } \Bigr) 
\sum_{d_a \ge 1, a\in \cB }^{\sum_{a\in \cB}d_a = 2|\cB|-2} \frac{(|\cB|-2)!}{\prod_{a\in \cB}(d_a-1)!}  
     \prod_{a\in \cB}  d_a! \; .
\]
The sum over $d_a$ is easily computed:
\[
 \sum_{ d_a \ge 1,a\in \cB }^{ \sum d_a  =2|\cB|-2 }  \prod_{a\in \cB} d_a= \binom{  3 |\cB| -3}{ |\cB|-2} \; ,
\]
as this sum is the coefficient of $x^{2 |\cB|-2}$ in the Taylor expansion of:
\[
 \Bigl( x \bigl( \frac{1}{1-x}\bigr)' \Bigr)^{ |\cB| } = \frac{ x^{ |\cB| } }{(1-x)^{2 |\cB| } } 
 = x^{|\cB| } \sum_{p\ge 0} \binom{ p+2|\cB|-1}{p} x^p \; .
\]
We thus reexpress the contribution of a block $\cB$ as:
\[
 \sum_{ \{j_a\}, a\in \cB } \;  |\lambda|^{2|\cB|-2} \sqrt{ (4|\cB| -4)!! } \;\; \frac{(3 |\cB| -3)!}{(2 |\cB|-1)!} \;\;
 \Bigl( \prod_{\genfrac{}{}{0pt}{}{a,b\in \cB}{a\neq b}} (1-\delta_{j_aj_b}) \Bigr)
   \Bigl( \prod_{a\in \cB} \frac{ 1}   { M^{ j_a-2 }   } \Bigr) \; .
\]

Note that the contribution of a Bosonic block only depends on the slice indices of its vertices and of the total number of its vertices. It is, in particular, 
completely insensitive to the Fermionic forest connecting the vertices. 
The hard core constraint inside each block imposes that the slice indices of its vertices are all different. It follows that:
\[
 \sum_{a\in \cB} j_a \ge j_{\min} + (j_{\min}+1) + \dots + (j_{\min} + |\cB|-1) =j_{\min}  |\cB|  + \frac{|\cB|(|\cB|-1)}{2}  \; ,
\]
because in the worst case one vertex of $\cB$ has slice index $j_a=j_{\min}$, another one has slice index 
$j_{a'}=j_{\min}+1$ and so on up to the last vertex which has slice index $j_{a''}= j_{\min}+|\cB|-1$. Therefore:
\[
  \sum_{a\in \cB} (j_a-2) = \sum_{a\in \cB} \frac{1}{2}j_a + \sum_{a\in \cB} (\frac{1}{2}j_a -2) 
  \ge  \sum_{a\in \cB} \frac{1}{2}j_a + \frac{j_{\min}-4}{2} |\cB| + \frac{|\cB|(|\cB|-1)}{4} \;,
\]
 and
 \beann
  && \sum_{ \{j_a\}, a\in \cB }  \Bigl( \prod_{\genfrac{}{}{0pt}{}{a,b\in \cB}{a\neq b}} (1-\delta_{j_aj_b}) \Bigr)
   \Bigl( \prod_{a\in \cB} \frac{ 1}   { M^{ j_a-2 }   } \Bigr) \crcr
   && \quad \le \Bigl( \sum_{j_a=j_{\min}}^{j_{\max}} \frac{1}{M^{\frac{j_a}{2}}} \Bigr)^{|\cB|}
   \frac{1}{M^{ \frac{j_{\min}-4}{2} |\cB| + \frac{|\cB|(|\cB|-1)}{4}   }} = 
   \frac{1}{M^{ \frac{|\cB|^2}{4}}} \Bigl[ \frac{1}{M^{j_{\min} -2- \frac14 -\frac12}} \; \frac{1}{M^{\frac12} -1}\Bigr]^{|\cB|} \; ,
 \eeann 
which is bounded by $M^{-\frac{|\cB|^2}{4}}$ for $j_{\min}\ge 3$ and for all $M>4 $. The essential point here is that this bound holds for any $j_{\max}$,
as the geometric series $\sum_{j_a=j_{\min}}^{j_{\max}}  M^{- \frac{j_a}{2}}  $ can be summed up to infinity. This is the reason why the MLVE allows to define the
ultraviolet limit of the model.
 
The sum over Fermionic forests can be computed as in subsection \ref{subsec:counting}. 
A forest over the blocks $\cB$ of the partition $\cP$ is a tree with $|\cP|-1$ edges and assigned coordination $D_{\cB}$ at the vertices plus a choice 
$|\cB|^{D_{\cB}}$ of the vertex inside each block to which the Fermionic edges hook:
\[
 \sum_{\cF_F} = (\sum_{\cB\in \cP} |\cB|)^{ |\cP| -2 } \prod_{\cB\in \cP} |\cB| \; .
\]
Grouping again together the partitions having $B_1$ blocks of size $1$, $B_2$ blocks of size $2$ and so on we get:
\beann
 && \Big{|}\log Z(\lambda, N) \Big{|}  \le  \sum_{n=1}^\infty \frac{1}{n!}   \sum_{B_1,\dots B_q,\dots \ge 0}^{\sum qB_q =n} 
     \frac{n!}{\prod_{q\ge 1} B_q! (q!)^{B_q} } 2^{\sum B_q -1}  n^{\sum B_q -2} \prod_{q\ge 1} q^{B_q}
    \crcr 
&& \qquad \qquad \times \prod_{q\ge 1} \Bigg[ 
  |\lambda|^{2q-2} \sqrt{ (4q -4)!! } \;\; \frac{(3 q-3)!}{(2 q-1)!} \;\; \frac{1}{M^{\frac{q^2}{4}}}
\Bigg]^{B_q} \; \;.
\eeann
We use the fact that $n^{ \sum B_q -2 } \le (\sum B_q)! \; e^{n}$ to obtain:
\beann
\sum_{B_1,\dots B_q,\dots \ge 0} \frac{ \bigl(\sum B_q \bigr)! }{\prod_q B_q! } \prod_{q\ge 1} 
  \Bigl[   \frac{ 2  \; e^{q} }{(q-1)!} \; |\lambda|^{2q-2} \sqrt{ (4q -4)!! } \;\; \frac{(3 q-3)!}{(2 q-1)!} \;\; \frac{1}{M^{\frac{q^2}{4}}}  \Bigr]^{B_q} \; .
\eeann
In order to establish the lemma, it is enough to prove that the combinatorial factor is bounded by:
 \[
  \frac{2}{(q-1)!} \sqrt{ (4q -4)!! }  \;\; \frac{(3 q-3)!}{(2 q-1)!} \le 3^{3q} e^{-q} q^q \; .
 \]
This bound obviously holds for $q=1$. For $q\ge 2$ we have:
\[
 \frac{2}{(q-1)!}  \sqrt{ (4q-4)!! }  \frac{(3 q-3)!}{(2 q-1)!} = \frac{ 4q^2 }{(3q)(3q-1)(3q-2) \sqrt{ (4q-1)(4q-3) }}  \sqrt{ (4q)!! } \frac{ (3q)!}{q! (2q)!}
 \le  \sqrt{ (4q)!! } \frac{ (3q)!}{q! (2q)!}
 \; ,
\]
which, using the upper and lower bounds on the factorial provided by the Stirling formula, is bounded by
\beann
 && \sqrt{ \frac{4q!}{2^{2q} (2q)! } }\frac{ (3q)!}{q! (2q)!} \le \sqrt{ \frac{1}{2^{2q} }  \frac{ (4q)^{4q +\frac12}  }{ (2q)^{2q+\frac12} } 
 \frac{e^{\frac{1}{12}}}{ e^{2q}} }  \frac{1 }{ \sqrt{ 2\pi}  }   \frac{(3q)^{3q+\frac12 } } { q^{q+\frac12} (2q)^{2q+\frac12} } e^{\frac{1}{12}} 
 \le 3^{3q} e^{-q} q^q \; .
\eeann 

\qed

Theorem \ref{thetheorem} now follows trivially from lemma \ref{lem:lastbound}. Indeed, the sum over $B_1,\dots B_q,\dots$ in equation \eqref{eq:lastbound} can be 
performed to yield:
\bea\label{eq:lastlastlast}
 \Big{|}\log Z(\lambda, N) \Big{|}  \le \sum_{B = 0}^{\infty} 
    \Bigl[ \sum_{q=1}^{\infty}  |\lambda|^{2q-2} \; 3^{3q}  \;  q^q \; \frac{1}{M^{\frac{q^2}{4}}}   \Bigr]^B \; .
\eea 
The sum over $q$ can always be rendered convergent by choosing $M$ large enough. Choosing for instance $M\ge 10^8$ ensures that
\[
  3^{3q} q^q \frac{1}{M^{\frac{q^2}{8}}} \le 1 \; , \qquad  \forall q\ge 1 \;,
\]
hence ensures that the sum over $q$ is bounded by 
\[
  \sum_{q=1}^{\infty} |\lambda|^{2q-2} \frac{1}{M^{\frac{q^2}{8}}} \le  \sum_{q=1}^{\infty} |\lambda|^{2q-2} \frac{1}{M^{\frac{q}{8}}} =\frac{1}{M^{1/8} - |\lambda|^2}
  \le 1,
\]
and this ensures in turn that the sum over $B$ in eq. \eqref{eq:lastlastlast} converges.

\subsection{Venturing in the complex plane}

In this last subsection we will extend our proof of convergence to a domain in the complex plane, and establish theorem \ref{thm:theorem2}.
The case of complex $\lambda$ follows the same lines as the one for real $\lambda$. The major difference arises 
at the level of lemma \ref{lem:boso}. For a complex coupling constant one needs to review the bound on the Bosonic 
Gaussian integral in a block.

We will treat the case when the Bosonic block has exactly one vertex (hence the tree inside the block has no edges) separately.
The Bosonic integral in this case is 
\[
 \int \frac{d\sigma}{\sqrt{2\pi}} e^{-\frac{1}{2} \sigma^2} \Bigr( \prod_{p\in I_{j_a}} \frac{1}{1-\imath \frac{\lambda \sigma }{p} }  
 e^{ -\imath  \frac{\lambda \sigma}  {p} } -1\Bigl) = -1+ 
 \int \frac{d\sigma}{\sqrt{2\pi}} e^{-\frac{1}{2} \sigma^2}   \prod_{p\in I_{j_a}} \frac{1}{1-\imath \frac{\lambda \sigma }{p} }  
 e^{ -\imath  \frac{\lambda \sigma}  {p} }  \; .
 \]
Reinstating the fields $\bar \phi_p,\phi_p$ we can rewrite this integral as
\beann
&& \int \Bigl(\prod_{p\in I_{j_a}} p\frac{ \bar \phi_p \phi_p }{ 2\pi \imath }\Bigr)e^{-\sum_{p\in I_{j_a} } p \bar \phi_p\phi_p  }
\Bigl[ e^{  - \frac{\lambda^2}{2} \Bigl[ \sum_{p\in I_{j_a} }   \bigl( \bar \phi_p \phi_p - \frac{1}{p} \bigr) \Bigr]^2 } -1 \Bigr] \crcr
&& = \int \Bigl(\prod_{p\in I_{j_a}} p\frac{ \bar \phi_p \phi_p }{ 2\pi \imath }\Bigr)e^{-\sum_{p\in I_{j_a} } p \bar \phi_p\phi_p  }
  \int_{0}^1 dt \Bigl(   - \frac{\lambda^2}{2} \Bigl[ \sum_{p\in I_{j_a} }   \bigl( \bar \phi_p \phi_p - \frac{1}{p} \bigr) \Bigr]^2 \Bigr)
e^{  - t \frac{\lambda^2}{2} \Bigl[ \sum_{p\in I_{j_a} }   \bigl( \bar \phi_p \phi_p - \frac{1}{p} \bigr) \Bigr]^2 } .
\eeann 
It is bounded for $\Re \lambda^2 \ge 0$ by:
\beann
\frac{ |\lambda|^2 }{2} \int \Bigl(\prod_{p\in I_{j_a}} p\frac{ \bar \phi_p \phi_p }{ 2\pi \imath }\Bigr)e^{-\sum_{p\in I_{j_a} } p \bar \phi_p\phi_p  }
   \Bigl[ \sum_{p\in I_{j_a} }   \bigl( \bar \phi_p \phi_p - \frac{1}{p} \bigr) \Bigr]^2 
   =\frac{ |\lambda|^2 }{2} \sum_{p \in I_{j_a}} \frac{1}{p^2} \le |\lambda|^2 \frac{1}{M^{j_a-2}} \; .
\eeann

For blocks having at least two vertices (hence the Bosonic tree has at least one edge), the case of complex $\lambda$ is more subtle.
Consider $\lambda = |\lambda|e^{\imath \gamma}$. If one follows the proof of lemma \ref{lem:boso}, one needs to establish a bound on
$\int d\nu \; \Big{|} e^{-V_{j_a}(\sigma^a_{j_a})} \Big{|}^4 $, which can be done using:
\[
\Big{|} e^{ -V_{j_a}(\sigma^a_{j_a}) } \Big{|} = \Big{|}\prod_{p\in I_{j_a}} \frac{1}{1-\imath \frac{\lambda \sigma^a_{j_a}}{p} } 
 e^{ -\imath  \frac{\lambda \sigma^a_{j_a}}{p} } \Big{|} \le \prod_{p\in I_{j_a}} \frac{1}{\cos \gamma} e^{|\lambda|\sin\gamma  \frac{ \sigma^a_{j_a}}{p} }
 \sim \frac{1}{(\cos\gamma)^{ M^{j_a} } } e^{|\lambda|\sin\gamma  \sum_{p\in I_{j_a}} \frac{ \sigma^a_{j_a}}{p} } \; .
\]
This bound is not optimal: the prefactors pile up and the block Bosonic Gaussian integral acquires a global prefactor 
$\frac{1} {(\cos\gamma)^{ \sum_{a\in \cB} M^{j_a}  } }  $ which will always beat the suppression $\frac{1}{M^{j_a}}$, and will spoil the 
summability over $j_a$. Therefore one needs to be careful and find a tighter bound on the Gaussian Bosonic integral. 

The solution comes from the remark that, while 
one needs to sum the slice attributions $j_a$ in a block (and each slice $j_a$ has $M^{j_a}$ indices), 
the number of vertices of the block is fixed. It is then much better to pay a large factor per vertex of the Bosonic block than to pay
one large factor per index in a slice. This can be done by turning the $\sigma$'s by a phase. 

Indeed, going back to the Bosonic Gaussian integral, 
\[
 \int d\nu  \; \Bigl( \prod_{a\in \cB}  e^{-V_{j_a} (\sigma^a_{j_a}) } \Bigr) 
 \Bigl( \prod_{a\in \cB} (\sigma^a_{j_a} )^{m^{(a)}_1} \Bigr) \Bigl( \prod_{a\in \cB} F_a( \sigma^a_{j_a} ) \Bigr) \; ,
\]
the Cauchy-Schwarz inequality should be applied after making the change of variables $\tau^a_{j_a} = e^{\imath \gamma} \sigma^a_{j_a}$, as this will 
give the same bounds as before on $ |  e^{-V_{j_a} (\sigma^a_{j_a}) }  |$ and $ | F_a( \sigma^a_{j_a} ) | $ (getting rid of the bad prefactor). However, 
when taking absolute values, the Gaussian integral over the $|\cB|$ distinct 
fields $\tau^a_{j_a}$ is not normalized anymore, and one needs to reestablish this normalization by a rescaling 
$\tilde \tau^a_{j_a} = \sqrt{\cos 2\gamma} \tau^a_{j_a} $. Taking into account that $\prod_{a\in \cB} (\sigma^a_{j_a})^{m_1^{(a)}}$ becomes
$ \prod_{a\in \cB} \frac{1}{\sqrt{\cos2\gamma}  } \tilde \tau^a_{j_a}  $, and the bound:
\[
  \sum_{\genfrac{}{}{0pt}{}{\forall a \in \cB} 
   {m^{(a)}_1,m^{(a)}_2, \dots \ge 0 }}^{\sum_{k\ge 1} km^{(a)}_k = d_a} \sqrt{(4 |\cB| -4)!!} \prod_{a\in \cB }  
   \Bigg[ d_a! \; |\lambda|^{d_a}   
     \frac{ |  \lambda|^{m^{(a)}_1   }  }{\prod_{k\ge 1}   m^{(a)}_k!  \; k  ^{m^{(a)}_k}  }
   \frac{1}{M^{j_a-2}}
    \Bigg] \; , 
\]
is replaced by:
\[
 \frac{1}{(\cos 2\gamma)^{\frac{|\cB|}{2}}} \sum_{\genfrac{}{}{0pt}{}{\forall a \in \cB} 
   {m^{(a)}_1,m^{(a)}_2, \dots \ge 0 }}^{\sum_{k\ge 1} km^{(a)}_k = d_a} \sqrt{(4 |\cB| -4)!!} \prod_{a\in \cB }  
   \Bigg[ d_a! \; |\lambda|^{d_a}   \frac{ |  \lambda|^{m^{(a)}_1   } }{ (\cos 2\gamma)^{ \frac{m^{(a)}_1 }{2} } }
     \frac{ 1 }{\prod_{k\ge 1}   m^{(a)}_k!  \; k  ^{m^{(a)}_k}  }
   \frac{1}{M^{j_a-2}}
    \Bigg] \; .
\]
In the domain $|\lambda|^2<\cos2\gamma$ in the complex plane we have:  
\[ 
  \frac{ |  \lambda|^{m^{(a)}_1   } }{ (\cos 2\gamma)^{ \frac{m^{(a)}_1 }{2} } }\le 1 \;,
\]
and lemma \ref{lem:boso} is replaced by
 \begin{lemma} For $\lambda = |\lambda| e^{\imath \gamma}$, $|\lambda|^2<\cos2\gamma $  the bosonic Gaussian integral in eq. \eqref{eq:bosogauss} 
 is bounded by \[
 |\lambda|^2 \frac{1}{M^{  j_a-2 } } \; ,
\]
for $|\cB|=1$ and for $|\cB|\ge 2$  it is bounded by
 \[
 \frac{1}{(\cos2\gamma)^{\frac{|\cB|}{2}}} \sqrt{(4|\cB|-4)!!} \; \; \Bigl(  \prod_{a\in \cB}  d_a! \; |\lambda|^{d_a} \frac{1}{M^{  j_a-2 } }  \Bigr) \; .
 \]
 \end{lemma}

 Everything else follows as in the real case, except that equation \ref{eq:lastlastlast} is replaced by  
 \beann
 \Big{|}\log Z(\lambda, N) \Big{|}  \le \sum_{B = 0}^{\infty} 
    \Bigl[ 
      \frac{ |\lambda|^2 }{M-1} +
    \sum_{q=2}^{\infty}  \frac{1}{ (\cos2\gamma)^{\frac{q}{2}} }|\lambda|^{2q-2} \; 3^{3q}  \;  q^q \; \frac{1}{M^{\frac{q^2}{4}}}   \Bigr]^B \; ,
\eeann 
which converges, for the same reasons as in the real case, in the domain $|\lambda|^2 <   (\cos2\gamma)$. This establishes theorem \ref{thm:theorem2}.

The domain $ |\lambda|^2 <   (\cos2\gamma) $ is relevant because, in terms of the original coupling constant of the model $g=\lambda^2$
it writes as 
\[
 \Re \frac{1}{g} \ge 1 \; ,
\]
that is our domain of convergence is exactly the Borel disk of center $\frac{1}{2}$ and radius $\frac{1}{2}$.

The theorem \ref{thm:theorem2} is the starting point to show that $\log Z(g=\lambda^2,N)$ is Borel summable in $g$ uniformly in $N$,
hence it is the Borel sum of its perturbative series in (renormalized) Feynman graphs.
In order to establish the full Borel summability one still needs to prove that after expanding in $g$ up to order $p$, the Taylor-Lagrange remainder term 
is bounded by $(gK)^p p!$. This follows straightforwardly as usual from bounding the integral 
formula for this remainder term with the techniques of this paper; the details are left to the reader.
 
\section{Conclusion}

In this exploratory paper we introduce a method to accommodate the subtleties of renormalization in the LVE expansion.
The convergence of the logarithm of the partition function defined in eq. \ref{eq:partitionfunction} is not surprising. The purpose of this 
paper is to introduce the method to obtain this result starting from the LVE expansion, that is starting from eq. \ref{eq:partitionfunctionintfield}.

The main feature of the MLVE we introduce in this paper is the hard core constraint on the slices of the vertices in each block. 
Such a hard core constraint is not surprising, as it is the fundamental feature of the usual Mayer expansion (which we recall in appendix \ref{app:Mayer}).
The MLVE is the right formalism to put this together with the LVE.

This hard core constraint is crucial. Indeed, the  $M^{-\frac{q^2}{4}}$ suppression in the sum over $q$ in eq. \eqref{eq:lastlastlast} comes from this hard
core constraint and is responsible for rendering it convergent. As is apparent from lemma \ref{lem:lastbound}, the combinatorial factor alone 
behaves like $q^q$, hence in the absence of a strong suppression at large $q$, the sum over $q$ would always diverge. This is exactly what goes wrong 
with the usual LVE expansion: the bad combinatorial factor $q^q$ coming from the proliferation of counterterms on trees with many leaves cannot be compensated. 
Curing this problem is the motivation of this work. 

The MLVE should now be applied to more interesting combinatorial quantum field theories, 
first superrenormalizable (eg such as the one defined in \cite{Carrozza:2012uv}), then renormalizable, such as the Grosse-Wukenhaar model 
\cite{Grosse:2004yu,Grosse:2009pa,Grosse:2012uv},
or the just renormalizable tensor models defined in \cite{BenGeloun:2011rc},  \cite{BenGeloun:2012pu} and \cite{Carrozza:2013wda}.

 \section*{Appendices}
 
\appendix 
 
\section{The Forest Formula}\label{app:forest}

A forest formula is a kind of Taylor expansion with integral remainder 
which expands a quantity depending on a symmetric positive $n$ by $n$ matrix. \emph{Positive} forest formulas, i..e. formulas in which the Taylor 
remainder integration paths always remain on the convex set of positive matrices are particularly interesting for constructive theory. A beautiful such positive forest formula which 
is symmetric under action of the permutation group on the $n$ points was discovered in \cite{BK} and developed with alternative proofs in \cite{AR1}\footnote{.
Non-symmetric versions appeared earlier in the constructive literature (see \cite{Rivasseau:2013tpa} for a recent reference).}.

Consider the vector space $S_n$ of  symmetric $n$ by $n$ matrices $X= \{X_{ij}\}, i,j = 1 ,\cdots, n$. It has dimension $n(n+1)/2$.
The set $PS_n$ of positive symmetric matrices whose diagonal coefficients are all equal to 1 and off-diagonal elements are 
between 0 and 1 is compact and convex. Symmetric matrices with diagonal elements equal to one and off-diagonal elements in $[0,1]^{n(n-1)/2}$
do not always belong to $PS_n$, as for instance the matrix $\begin{pmatrix} 1 &1&0 \cr 1 &1&1 \cr 0 &1&1 \end{pmatrix}$ is not positive. But
$PS_n$ has an $n(n-1)/2$ dimensional interior, hence any matrix $X\in PS_n$ can be parametrized by $n(n-1)/2$ elements $X_\ell$, 
where $\ell$ runs over the edges of the complete graph $K_n$. 

$PS_n$ contains as particularly interesting elements the block matrices $X^\Pi$ for any partition $\Pi$ of $V$. Any such partition
divides the entries of any symmetric $n$ by $n$ matrix into same-block and trans-block entries. The
block matrix $X^\Pi$ has entries $X_{ij}^\Pi =1 $ if $i$ and $j$ belong to the same block of the partition, and 0 otherwise. 
Two extremal cases are the identity matrix  
$\mathbb{I}$, corresponding to $X^{sing}$, that is to the maximal partition made of all singletons, and
the matrix $\bbone$ with all entries equal to one, corresponding to $X^{[1,\cdots, n]}$, that is to the minimal partition made of a single block.
The forest formula can be considered as interpolating between these two extremal points.

Let us consider a function $f$ defined and smooth in the interior of $PS_n$ with continuous extensions (together with all their derivatives) to $PS_n$ itself.
The precise statement is

\begin{theorem}[The Forest Formula]
\bea  f( {\bbone }) =      \sum_{\cF} \int dw_\cF  \;  \partial_\cF  f \, [ X^\cF (w_\cF) ] \label{bkar}
\eea
where
\begin{itemize}

\item The sum over $\cF$ is over forests over $n$ labeled vertices $i = 1, \cdots , n$, including the empty forest with no edge. Such forests
are exactly the acyclic edge-subgraphs of the complete graph $K_n$.

\item  $\int dw_\cF$ means integration from 0 to 1 over one parameter for each forest edge: $\int dw_\cF  = \prod_{\ell\in \cF}  \int_0^1 dw_\ell  $.
There is no integration for the empty forest since by convention an empty product is 1. A generic integration point $w_\cF$
is therefore made of $\vert \cF \vert$ parameters $w_\ell \in [0,1]$, one for each $\ell \in \cF$.

\item  $ \partial_\cF = \prod_{\ell\in \cF} \partial_\ell  $ means a product of first order partial derivatives with respect to 
the variables $X_{\ell}$ corresponding to the edges of $\cF$. Again there is no such derivatives for the empty forest since by convention an empty product is 1.

\item $X^\cF (w_\cF)$ is defined by $X^\cF_{ii} (w_\cF )= 1$ $\forall i$, and for $i \ne j$
by $X^\cF_{ij} (w_\cF)$ being the infimum of the $w_\ell$ parameters for $\ell$
in the unique path $P^\cF_{i \to j}$ from $i$ to $j$ in $\cF$, when such a path exists. If no such path exists, which means that $i$ and $j$ belong to different connected components with respect to the forest 
$\cF$, then by definition $X^\cF_{ij} (w_\cF) = 0$.

\item The symmetric $n$ by $n$ matrix $X^\cF (w_\cF)$ defined in this way is positive, hence
belongs to $PS_n$, for any value of $w_\cF$.

\end{itemize}
\end{theorem}
\noindent
Since $X^\emptyset = \mathbb{I}$, the empty forest term in \eqref{bkar} is $f(\mathbb{I})$, hence \eqref{bkar} indeed interpolates $f$ between $\bbone$ and $\mathbb{I}$, staying on $PS_n$ as announced.

We shall not repeat the proof, as it has been detailed in \cite{BK,AR1}. We prefer to recall a useful corollary of this theorem which expands Gaussian integrals over replicas.
Consider indeed a Gaussian measure $d\mu_C$ of covariance $C_{pq}$ on a vector variable $\vec \tau$ with $N$
components $\tau_p$. To study approximate factorization properties of the integral of a product of $n$ functions of the variable $\vec \tau$ 
it is useful to first rewrite this integral using a replica trick. It 
means writing the integral over $n$ identical replicas $\vec \tau_i$ for $i=1, \cdots , n$ with components $\tau_{p,i}$, with the 
perfectly well-defined measure with covariance $[C\otimes \bbone]_{p,i ; q,j} = C_{pq}$ :
\bee  \int d\mu_C (\vec \tau)    \prod_{i=1}^n  f_i(\vec \tau)  =   \int d\mu_{C\otimes \bbone} (\vec \tau_i)    \prod_{i=1}^n  f_i(\vec \tau_i) .
\ee
Applying the forest formula we obtain the following corollary

\begin{cor}
\bea I= \int d\mu_C (\vec \tau)    \prod_{i=1}^n  f_i(\vec \tau)  =  \sum_{\cF} \int dw_\cF  \int d\mu_{C\otimes X^\cF  (w_\cF)} (\vec \tau_i)      
\;  \partial^C_\cF  \prod_{i=1}^n  f_i(\vec \tau_i)     \label{bkar5}
\eea
where $\partial^C_\cF$ means $\prod_{\ell =(i,j)\in \cF} \frac{\partial}{\partial \tau_{p,i}} C_{pq} \frac{\partial}{\partial \tau_{q,j}} $ 
and we use Einstein's convention for the sums over $p$ and $q$. \end{cor}

\prf 
Using the general Wick theorem in $n$ variables $x_i$ 
\bee \int d\mu_C  f (x) =  e^{  \frac{\partial}{\partial \tau_i} C_{ij}  \frac{\partial}{\partial \tau_j}} f \vert_{\tau=0}
\ee
leads to the corollary in a rather straightforward way. \qed

\section{The Mayer Expansion}\label{app:Mayer}

The Mayer expansion \cite{mayer} is a statistical mechanics method to compute the free energy of gases of polymers 
which are subsets of a ``monomer" set $M$ with interactions. In constructive theory context it is applied to factorize the
abstract hardcore interactions resulting from a cluster expansion \cite{brydgesfed}.
It is then equivalent to the Fermionic part of the MLVE defined above, but is not usually formulated
in terms of Grassmann variables.

Suppose we consider a partition function which is a sum of polymer activities with hardcore contraints
\bea \label{mayer0}  Z(\lambda, N) &=&  \sum_{ \genfrac{}{}{0pt}{}{P_1, \cdots, P_n}{ P_i \subset M, P_i \cap P_j = \emptyset } } \prod_{i=1}^n A(P_i) . \eea
Using replicas 
\bea \label{mayer1}  Z(\lambda, N) &=&  \sum_{n=0}^{\infty}  \frac{1}{n!}  \sum_{P_1, \cdots, P_n } \prod_{i=1}^n A(P_i) \prod_{1 \le i < j \le n} \epsilon_{ij} ,\eea
where $\epsilon_{ij} = 0 $ if $P_i \cap P_j = \emptyset$. Defining $\epsilon_{ij} = 1 + \eta_{ij} = 1+x_{ij} \eta_{ij} \vert_{x_{ij} = 1}$, and applying the forest formula
allows to compute the free energy as
\bea \label{mayer2}  \log Z(\lambda, N) &=&  \sum_{n=0}^{\infty} 
\frac{1}{n!} \sum_{\cT} \sum_{P_1, \cdots, P_n } \prod_{i=1}^n A(P_i) \epsilon^\cT ,
\eea
where the sum is over spanning trees $\cT$ over $[1, \cdots ,n]$ and
\bee
\epsilon^\cT =  \bigg\{ \prod_{\ell\in \cT}    \big[ \int_0^1 dw_\ell \big]    \eta_\ell \bigg\}\prod_{\ell \not\in \cT} \big[ 1 +  \eta_{\ell}  X^\cF_\ell (\{ w\}) \big],
\ee
with $ X^\cF_\ell (\{ w\}) $ defined as in the previous section.  is bounded by 1.

Since $\epsilon^\cT \le 1$, summation over the polymer shapes can proceed from the leaves towards the root of the Mayer tree in a standard way.
Under the typical condition 
\bee  \sum_{P \supset p_0} \vert  A (P) \vert e^{\vert P \vert } < 1 ,
\ee
where $p_0$ is a root monomer in $M$,
the Mayer expansion is absolutely convergent  \cite{brydges1}.


\begin{thebibliography}{99}


\bibitem{GJ} J. Glimm and A. M. Jaffe, ``Quantum Physics. A Functional Integral Point Of View,Ó New York, Springer (1987). 

\bibitem{Riv}
V. Rivasseau, ``From perturbative to constructive renormalization,Ó
Princeton University Press (1991). 


\bibitem{Rivasseau:2013ova} 
  V.~Rivasseau and Z.~Wang,
  ``How to Resum Feynman Graphs,''
  arXiv:1304.5913 [math-ph].

  
\bibitem{BK} D. Brydges and T. Kennedy,
Mayer expansions and the Hamilton-Jacobi equation, Journal of
Statistical Physics, {\bf 48}, 19 (1987).

\bibitem{AR1}
 A.~Abdesselam and V.~Rivasseau,
 ``Trees, forests and jungles: A botanical garden for cluster expansions,''
 arXiv:hep-th/9409094.


\bibitem{Rivasseau:2007fr} 
  V.~Rivasseau,
  ``Constructive Matrix Theory,''
  JHEP {\bf 0709}, 008 (2007),
  arXiv:0706.1224. 

\bibitem{RW}{V.~Rivasseau and Zhituo Wang,
Loop Vertex Expansion for $\phi^{2k}$ Theory in Zero Dimension, arXiv:1003.1037, J.\ Math.\ Phys.\  {\bf 51} (2010) 092304}

\bibitem{MR1}
  J.~Magnen and V.~Rivasseau,
  ``Constructive $\phi^4$ field theory without tears,''
  Annales Henri Poincar\'e {\bf 9} (2008) 403
  [arXiv:0706.2457 [math-ph]].

\bibitem{Gurau:2011xp} 
  R.~Gurau and J.~P.~Ryan,
  ``Colored Tensor Models - a review,''
  SIGMA {\bf 8}, 020 (2012)
  [arXiv:1109.4812 [hep-th]].
  
\bibitem{Rivasseau:2013uca} 
  V.~Rivasseau,
  ``The Tensor Track, III,''
  arXiv:1311.1461 [hep-th].

\bibitem{Di Francesco:1993nw}
  P.~Di Francesco, P.~H.~Ginsparg and J.~Zinn-Justin,
  ``2-D Gravity and random matrices,''
  Phys.\ Rept.\  {\bf 254}, 1 (1995)
  arXiv:hep-th/9306153.
 
  \bibitem{sefu1}
  J.~Magnen, K.~Noui, V.~Rivasseau and M.~Smerlak,
  ``Scaling behavior of three-dimensional group field theory,''
  Class.\ Quant.\ Grav.\  {\bf 26}, 185012 (2009),
  arXiv:0906.5477. 

\bibitem{Gurau:2013pca} 
  R.~Gurau,
  ``The $1/N$ Expansion of Tensor Models Beyond Perturbation Theory,''
  arXiv:1304.2666.
  
\bibitem{BenGeloun:2011rc}
  J.~Ben Geloun and V.~Rivasseau,
``A Renormalizable 4-Dimensional Tensor Field Theory,''
  arXiv:1111.4997. 



\bibitem{Carrozza:2012uv} 
  S.~Carrozza, D.~Oriti and V.~Rivasseau,
  ``Renormalization of Tensorial Group Field Theories: Abelian U(1) Models in Four Dimensions,''
  arXiv:1207.6734. 

\bibitem{Carrozza:2013wda} 
  S.~Carrozza, D.~Oriti and V.~Rivasseau,
  ``Renormalization of an SU(2) Tensorial Group Field Theory in Three Dimensions,''
  arXiv:1303.6772. 
  
\bibitem{Geloun:2013saa} 
  J.~Ben Geloun,
  ``Renormalizable Models in Rank $d\geq 2$ Tensorial Group Field Theory,''
  arXiv:1306.1201. 

\bibitem{Disertori:2006nq} 
  M.~Disertori, R.~Gurau, J.~Magnen and V.~Rivasseau,
  ``Vanishing of Beta Function of Non Commutative Phi**4(4) Theory to all orders,''
  Phys.\ Lett.\ B {\bf 649}, 95 (2007),
  hep-th/0612251.
  
\bibitem{BenGeloun:2012pu} 
  J.~Ben Geloun and D.~O.~Samary,
  ``3D Tensor Field Theory: Renormalization and One-loop $\beta$-functions,''
  Annales Henri Poincar\'e {\bf 14}, 1599 (2013),
  arXiv:1201.0176. 
  



\bibitem{BenGeloun:2012yk} 
  J.~Ben Geloun,
  ``Two and four-loop $\beta$-functions of rank 4 renormalizable tensor field theories,''
  Class.\ Quant.\ Grav.\  {\bf 29}, 235011 (2012),
  arXiv:1205.5513. 
  
  \bibitem{Simon}
  B.~Simon,
  ``The $P(\Phi)_2$ Euclidean (Quantum) Field Theory,''
{\it  Princeton University Press,\ princeton 1974, 392 P.(Princeton
Series In Physics)}

  
\bibitem{nelson}
E. Nelson, ``A quartic interaction in two dimensions", Mathematical Theory of Elementary Particles, Cambridge, M.I.T. Press, 1965, pp. 69Ð73.  

\bibitem{abdesselam} A. Abdesselam, V. Rivasseau,
``Explicit Fermionic Tree Expansions", Letters in Mathematical Physics, Vol.44, 77-88, 1998.

\bibitem{disertori} M. Disertori and V. Rivasseau, ``Continuous Constructive Fermionic Renormalization",
Annales Henri Poincar\'e, Vol. 1,   1-57 ( 2000)
 
\bibitem{Grosse:2004yu}
  H.~Grosse and R.~Wulkenhaar,
``Renormalisation of phi**4 theory on noncommutative R**4 in the matrix base,''
  Commun.\ Math.\ Phys.\  {\bf 256}, 305 (2005),
   arXiv:hep-th/0401128.
  
\bibitem{Grosse:2009pa} 
  H.~Grosse and R.~Wulkenhaar,
  ``Progress in solving a noncommutative quantum field theory in four dimensions,''
  arXiv:0909.1389. 
  
\bibitem{Grosse:2012uv} 
  H.~Grosse and R.~Wulkenhaar,
  ``Self-dual noncommutative $\phi^4$-theory in four dimensions is a non-perturbatively solvable and non-trivial quantum field theory,''
  arXiv:1205.0465. 
 

 
\bibitem{Rivasseau:2013tpa} 
  V.~Rivasseau and A.~Tanasa,
  ``Generalized constructive tree weights,''
  arXiv:1310.2424 [math-ph].
  
\bibitem{mayer}
Mayer, Joseph E.; Montroll, Elliott ``Molecular distributions", J. Chem. Phys. 9: 2Ð16, (1941).

\bibitem{brydgesfed}
D. Brydges and P. Federbush ``A New Form of the Mayer Expansion in Classical Statistical Mechanics",
Journ. Math.Phys. 19 (1978) 2064.

\bibitem{brydges1} D. Brydges
``A short course on Cluster Expansions", Les Houches, Session XLIII, 1984
K. Osterwalder and R. Stora, eds. Elsevier 1986.



\end{thebibliography}
\end{document}